\documentclass[11pt,a4paper]{article}

\usepackage{jheppub} 

\usepackage{graphicx} 
\usepackage{physics}  
\usepackage{amsmath}
\usepackage{amsthm}
\usepackage{amsfonts}
\usepackage{xspace}
\usepackage{subcaption}
\usepackage{ifthen}
\usepackage{tikz}
\usepackage{hyperref}
\usepackage{color}
\usepackage{subcaption}

\newcommand{\pd}{\partial} 
\newcommand{\sn}{\mathrm{sn}}
\newcommand{\cn}{\mathrm{cn}}
\newcommand{\dn}{\mathrm{dn}}

\newcommand{\revise}[1]{#1}

\numberwithin{equation}{section}

\title{Analytic Approach for Computation of Topological Number of Integrable Vortex on Torus}

\author{Kaoru Miyamoto}
\author{and Atsushi Nakamula}

\affiliation{
  Department of Physics, School of Science, Kitasato University\\
  Sagamihara 252-0373, Japan
}

\emailAdd{miyamoto.kaoru@st.kitasato-u.ac.jp, nakamula@sci.kitasato-u.ac.jp, nakamula00@gmail.com}

\abstract{\revise{Detailed structures of vortices on a torus are discovered by performing} an analytic method to calculate the vortex number.
We focus on analytic vortex solutions to the Chern-Simons-Higgs theory, whose governing equation is the so-called Jackiw-Pi equation.
The equation is one of the integrable vortex equations and is reduced to Liouville's equation.
The requirement of continuity of the Higgs field strongly restricts the characteristics and the fundamental domain of the vortices.
Also considered are the decompactification limits of the vortices on a torus, in which ``flux loss" phenomena occasionally occur.}

\begin{document}

\maketitle

\section{Introduction}

Vortices are ubiquitous structures in various scales of nature.
They typically demonstrate a topologically non-trivial configuration of fields both in quantum and classical dynamics \cite{Abrikosov:1956sx, Nielsen:1973cs,Manton:2004tk}. 
We define the vortices here as localised static solutions to a gauge theory coupled with a matter, or a Higgs field.   
A characteristic example of the vortices is magnetic flux in type-II superconductors, which appeared as topological solitons in the Ginzburg-Landau(GL) model, namely, the static energy functional of a 2+1 dimensional Abelian-Higgs model.
Those vortices in the GL model emerge as defects of field configuration with a spontaneous phase transition of the system, in which the ordinary Maxwell electromagnetism governs the dynamics of vortices described by an Abelian gauge field.
In 2+1 dimensions or 3 spatial dimensions, another ``dynamics" for gauge fields is possible: one can incorporate the Chern-Simons three-form, which brings topological degrees of freedom.
The physical significance of the Chern-Simons term stems from the fact that a theory including it may serve as an effective theory of the quantum Hall effect \cite{Tong:2016kpv,Klitzing2020}. 
Vortex solutions may notably exist in the Maxwell-Chern-Simons-matter theories and the pure Chern-Simons-matter theories \cite{Horvathy:2008hd}. 
In this paper, we consider the Jackiw-Pi vortex equation, which governs the non-relativistic Chern-Simons-matter theory in a static case and its vortex solutions.

In ordinary Abelian-Higgs vortices such as in type-II superconductors, flux of a gauge field, i.e., magnetic flux, is concentrated at zeroes of the Higgs field.
However, another kind of vortices exists for which the magnetic flux is excluded from the Higgs zero, sometimes referred to as exotic vortices \cite{Walton:2021dgi}.
The Jackiw-Pi vortices belong to the latter case.

As with ordinary field theories, field equations governing vortices are also second-order differential equations.
However, situations in which they become first-order equations exist if the coupling constants obey critical relations, i.e., the BPS (Bogomolnyi-Prasad-Sommerfield) limits.
In such cases, the system of first-order BPS equations is equivalent to the celebrated Liouville equation, a second-order solvable differential equation.
In \cite{Manton:2016waw}, Manton shows that five distinct cases of such integrable vortex equations exist according to curvature of the background surfaces on which vortices live. 
Among the five equations, the Taubes equation \cite{Taubes:1979tm}, the Ambj\o rn-Olesen equation \cite{Ambjorn:1988fx}, and the Bradlow equation \cite{Bradlow:1990ir} are defined on a hyperbolic surface $\mathbb{H}^2$,
while the Popov equation \cite{Popov:2007ms,Popov:2008gw} is defined on a sphere $S^2$.
The integrable vortex equation on a plane $\mathbb{R}^2$ is the Jackiw-Pi equation \cite{Jackiw:1990aw,Jackiw:1990tz} considered in the present paper.
There exist some geometrical interpretations behind those integrable vortex equations.
In \cite{Baptista:2012tx}, the Higgs fields of vortices are explained as conformal factors of a metric of the constant curvature surface with isolated singularities.
In \cite{Contatto:2017alh}, it is interpreted that these integrable equations can be reduced from four-dimensional Yang-Mills theories, and the relation between vortices and a flat non-Abelian connection in three-dimensions is shown in \cite{Ross:2021afj}.
In addition, higher-order generalizations in terms of the ``vortex polynomials" to those integrable vortex equations are considered in \cite{Gudnason:2022hxu}, which includes equations from Chern-Simons theories.

Although the integrable vortices except for the Popov vortices are defined on non-compact surfaces, they may also live on compact surfaces of constant curvature, i.e., surfaces of genus $g\geq1$.
For those vortices, periodicity of the background surfaces strongly restricts solution spaces to the vortex equations.
For the cases of $g=1$, the Jackiw-Pi vortices on a torus are considered earlier in \cite{Olesen:1991dg} and reconsidered in \cite{Akerblom:2009ev}, in which the elliptic functions describe vortices.
\revise{In addition, the relativistic Chern-Simons-Higgs vortices also exist on a torus \cite{Caffarelli1995}.}
For the cases of $g=2$, a special solution to the Taubes equation is constructed with the Schwarzian triangular functions \cite{Manton:2016waw}.
Along these lines, there exist several discussions regarding a vortex number on a compact manifold for the Popov vortex \cite{Popov:2012av,Manton:2012fv} and the Taubes vortex \cite{Baptista:2012tx}.
However, a comprehensive understanding of \revise{analytical aspects of} vortices on compact surfaces has not been \revise{accomplished.}
In this paper, we will \revise{perform explicitly} an analytical \revise{calculation} for determining a vortex number on a torus, namely the first Chern number, and aim to elucidate the vortices on compact surfaces in detail.

Another topic of solitonic objects on compact spaces such as torus is the possibility of having twisted periodic conditions, which leads to so-called fractionally charged solitons, e.g., \cite{Bruckmann:2007zh}.
Those kinds of objects are one of the key research interests for the physics of confinement \cite{tHooft:1981nnx,Gonzalez-Arroyo:2019wpu,Anber:2023sjn}, however, we do not consider the twisted periodic conditions here and remain them as a subject of future research.

The study of topological objects such as vortices, skyrmions, etc., is actively performed and still underway. 
Although integrability of such systems plays a significant role, soliton-like phenomena also appear in the systems without integrability, e.g., \cite{Koike:2022gfq}.
These are important interdisciplinary studies of physics, mathematics, and other areas of science.
In particular, these solitonic objects on spaces with periodicity, which may be interpreted as compact spaces, would provide an important contribution to condensed matter physics as well as mathematical physics.
In this context, a possibility of interpreting water waves as topological solitons has been proposed in \cite{Smirnova:2023}.
This direction would open a new window into the research for topological solitons.
It is important to clarify the meaning of integrability in nature.

This paper is organized as follows.
In the next section, we define the models deriving the Jackiw-Pi equation and give an outline of the integrable vortex equations.
In section 3, we demonstrate
the analytic calculation of a vortex number on compact surfaces and analyse typical cases with specific examples.
In section 4, the large period limits of vortex solutions are considered on a torus.
In the final section, we give conclusions and discussions.

\section{The Jackiw-Pi equation}

In this section, we briefly review the Jackiw-Pi vortex equation emerging from two different field theories.
In section \ref{sec:manton-five-vortex}, the five integrable vortex equations derived systematically from the Abelian Higgs model on constant curvature surfaces are discussed.
These equations are classified by Manton \cite{Manton:2016waw}, and the Jackiw-Pi equation is one of them.
In section \ref{sec:Non-relativ-CS}, the Jackiw-Pi equation as a static equation of the non-relativistic Chern-Simons-matter theory is outlined to justify a stability of the Jackiw-Pi vortices.
In section 2.3, it is noted that general solutions to the Jackiw-Pi equation are given in terms of a meromorphic function defined on a flat plane $\mathbb{C}$ or a torus $T^2$ in both theories.

\subsection{Integrable vortices}\label{sec:manton-five-vortex}

\revise{
To introduce the Jackiw-Pi equation, let us summarise Manton's paper\cite{Manton:2016waw}.
 Consider the Abelian Higgs model on a two-dimensional surface with the critical coupling constants\cite{Manton:2004tk}.
}
Let $M_0$ be a surface with conformal metric $ds_{0}^2=\Omega_0(dx^2+dy^2)$, where the conformal factor $\Omega_0$ is a function of $x$ and $y$. 
The static energy functional $E$ of such the Abelian Higgs model on $M_0$ is
\revise{
\begin{align}
    E = \int_{M_0} \qty{ -\frac{4}{\Omega_0^2}F_{z\bar{z}}^2 - \frac{4C}{\Omega_0}\qty(\abs{D_{z} \phi}^2 + \abs{D_{\bar{z}} \phi}^2) + \qty(-C_0+C\abs{\phi}^2)^2 }\frac{i\Omega_0}{2} dz \wedge d\bar{z},
\end{align}
}
where $z=x+iy$, $F_{z\bar{z}}=\pd_za_{\bar{z}}-\pd_{\bar{z}}a_z$ is a field strength, \revise{$D_z=\pd_z-ia_z$} and $D_{\bar{z}}=\pd_{\bar{z}}-ia_{\bar{z}}$ are covariant derivatives with respect to gauge potentials $a_z$ and $a_{\bar z}$, and $\phi$ is a complex Higgs field.
The real constants $C_0$ and $C$ will be specified later.
Applying the Bogomolny completion to this energy functional, we obtain the following formula,
\revise{
\begin{align}\label{eq:Bogomlny-completion}
    E = \int_{M_0} \qty{
        \qty(-\frac{2i}{\Omega_0}F_{z\bar{z}} + C_0 - C\abs{\phi}^2)^2
    -\frac{8C}{\Omega_0}\abs{D_{\bar{z}}\phi}^2
    }\frac{i\Omega_0}{2} dz \wedge d\bar{z}
    -4\pi C_0 N,
\end{align}
}
where 
\revise{\begin{align}\label{eq:first Chern number}
    N:=\frac{1}{2\pi}\int_{M_0} F_{xy} dx\wedge dy=\frac{1}{2\pi}\int_{M_0} F_{z\bar{z}} dz\wedge d\bar{z},
\end{align}}
is the first Chern number taking an integer value.
\revise{This} integer \eqref{eq:first Chern number} is interpreted as \revise{a vortex number on $M_0$} considering with multiplicity.
\revise{$N$ is also} a topological invariant because it is independent of a metric of $M_0$.
From this formula, if the following Bogomolny equations,
\begin{align}\label{eq:Bogomlny-eq}
    D_{\bar{z}}\phi = 0, \quad -\frac{2i}{\Omega_0}F_{z\bar{z}} = -C_0 + C\abs{\phi}^2,
\end{align}
are satisfied, we find the energy is bounded below $E\geq -4\pi C_0 N$ if $C\leq 0$, thus the field configurations satisfying \eqref{eq:Bogomlny-eq} are stable.
\revise{However, it is not the case when $C>0$, and we will avoid the instability from another point of view in the next section.}

Eliminating $F_{z\bar{z}}$ by using the first equation of \eqref{eq:Bogomlny-eq}, we obtain the following equation,
\begin{align}\label{eq:vortex-eq-general}
    \pd\overline{\pd} \log |\phi|^2 = \frac{\Omega_0}{2}\qty(C_0 - C\abs{\phi}^2),
\end{align}
where we have defined $\partial:=\partial_z$ and $\overline{\partial}:=\partial_{\overline{z}}$.
We refer to the equation \eqref{eq:vortex-eq-general} as the generalised vortex equation.


We note that values of the constants $C_0$ and $C$ can be normalised as $-1,0$ or $1$ by rescaling the metric and $\abs{\phi}$.
In ref.\cite{Manton:2016waw}, Manton argued that there are nine possible ways to choose $C_0$ and $C$, but the four cases of them are invalid:
The right-hand-side of \eqref{eq:vortex-eq-general} must be positive since the left-hand side of \eqref{eq:vortex-eq-general} is the magnetic field $F_{z\bar{z}}$ whose integral gives a positive topological number $N$.
Therefore the remaining five cases $(C_0,C)=(-1,-1),\ (-1,0),\ (-1,1),\ (0,1),\ \text{and}\ (1,1)$ are acceptable.

\revise{According to the geometrical interpretation of vortices \cite{Baptista:2012tx}, the general solution to \eqref{eq:vortex-eq-general} can be obtained as follows:}
\begin{align}\label{eq:solution-to-general}
    \abs{\phi}^2 = \frac{\left(1+C_0\abs{z}^2\right)^2}{\left(1+C\abs{f}^2\right)^2}\abs{\dv{f}{z}}^2,
\end{align}
\revise{where $f$ is an arbitrary meromorphic function on $M_0$.}

\revise{The Jackiw-Pi equation can be introduced as a special case of the generalized vortex equation \eqref{eq:vortex-eq-general},} which is the integrable vortex equation defined on a flat Euclidean plane.
\revise{To ensure the background surface becomes a flat plane, we set the constant $C_0=0$}
and the conformal factor $\Omega_0=4$.
From Manton's classification, the case of $(C_0,C)=(0,1)$ is the only possible case.
\revise{Hence,} the target surface is $S^2$, and the Jackiw-Pi equation takes the form
\revise{\begin{align}
    D_{\bar{z}}\phi = 0, \quad -2iF_{z\bar{z}} = 4\abs{\phi}^2, \label{eq:JP_equation_pre}
\end{align}}
\revise{or by combining them,}
\begin{align}
    \pd\overline{\pd} \log |\phi|^2 = -2\abs{\phi}^2.
    \label{eq:JP equation phi}
\end{align}
\revise{This is Liouville's equation exactly, which is one of the typical integrable equations.}
The solution to Liouville's, or the Jackiw-Pi equation is therefore,
\begin{align}\label{eq:solution-JP}
    \abs{\phi(z,\bar{z})}^2 = \frac{1}{(1+\abs{f}^2)^2}\abs{\dv{f}{z}}^2,
\end{align}
where $f$ is a meromorphic function on $\mathbb{C}\simeq \mathbb{R}^2$.
The meromorphicity of $f$ is required by the target surface being $S^2$.

\subsection{Non-relativistic Chern-Simons-matter theory and the Jackiw-Pi equation}\label{sec:Non-relativ-CS}

In this section, we review that the Jackiw-Pi equation can be derived from the $2+1$ dimensional non-relativistic Chern-Simons-matter theory \cite{Horvathy:2008hd}.
In contrast to the case of the Abelian Higgs model with the constant $C>0$, vortices are stable in this theory despite being governed by the same Jackiw-Pi equation.

\revise{Consider} the following Lagrangian density,
\begin{align}\label{eq:Chern-Simons-Lagrangian}
    \mathcal{L}_{JP} = i\Psi^{*}D_0\Psi - \frac{1}{2m}\abs{\vec{D}\Psi}^2 +\frac{\lambda}{2}\abs{\Psi}^4 + \frac{\kappa}{2}\epsilon^{\alpha\beta\gamma} A_{\alpha}F_{\beta \gamma},
\end{align}
    where $\Psi$ is a complex scalar field, $D_{\mu}=\pd_{\mu}-iqA_{\mu}$ is a covariant derivative with an Abelian gauge field $A_{\mu}$, $\vec{D}=(D_{1},D_{2})$, and $F_{\beta \gamma}$ is a field strength.
The metric and the complete anti-symmetric tensor are  defined as $g_{\mu\nu}=\mathrm{diag.}(-1,1,1)$ and $\epsilon^{012}=1$, respectively.

\revise{Let us focus on a static case of this theory.
The Euler-Lagrange equation with a static condition $\partial_0\Psi=\partial_0A_i=0$ is
\begin{align}\label{eq:static ansatz}
    - \frac{1}{2m}D_\mp D_\pm\Psi-\left(\lambda\mp\frac{q^2}{2m\kappa}\right)|\Psi|^2\Psi+qA_0\Psi&=0\\
    F_{12} + \frac{q}{\kappa}\abs{\Psi}^2 &= 0, \label{eq: F12}.
\end{align}
where $D_{\pm}=(D_1\pm iD_2)/2$,}
and $F_{12}$ is identical to $F_{xy}$ in \eqref{eq:first Chern number}.
Taking a gauge 
\begin{align}
    A_0=\pm\frac{q}{2m\kappa}\abs{\Psi}^2,
\end{align}
\revise{the equation} \eqref{eq:static ansatz} becomes
\revise{\begin{align}\label{eq:static ansatz with gauge fixing}
    - \frac{1}{2m}D_\mp D_\pm \Psi-\left(\lambda\mp\frac{q^2}{m\kappa}\right)|\Psi|^2\Psi=0.
\end{align}}
This can be solved by the following ``self-dual" or "anti-self-dual" ansatz
\begin{align}
    \revise{D_\pm\Psi=0,}\label{eq:SD equation}\\
    \lambda = \pm\frac{q^2}{m\abs{\kappa}}.
\end{align}
We refer to $\lambda=q^2/m\abs{\kappa}$ as the self-dual coupling, and $\lambda$ with opposite sign as the anti-self-dual coupling.
The defining equation for the scalar fields \eqref{eq:SD equation} is first order in its derivatives, so we refer to it as the BPS equation.
From the BPS equation \eqref{eq:SD equation} with \eqref{eq: F12}, we obtain the Jackiw-Pi equation again,
\begin{align}\label{eq:JP equation}
    \partial\bar{\partial}\log\abs{\Psi}^2=-\frac{1}{2}\abs{\Psi}^2,
\end{align}
where we have chosen $q=1$ and \revise{$\abs{\kappa}=1.$}
This is equivalent to \eqref{eq:JP equation phi} by setting $\Psi = 2\phi$.

The static energy with \revise{the} ansatz turns out to be
\revise{
\begin{align}
    E_{JP} = \int \qty(\frac{1}{2m}\abs{D_{\pm}\Psi}^2 - \frac{1}{2}\left(\lambda \mp \frac{q^2}{m\kappa}\right)\abs{\Psi}^4)d^2x \Bigg|_{\lambda=\pm\frac{q^2}{m\kappa}} = \int \frac{1}{2m}\abs{D_{\pm}\Psi}^2d^2x,
\end{align}
}
\revise{thus the static energy takes the minimum value $E_{JP}=0$ if $\Psi$ is a solution to the BPS equation.}
\revise{Hence, the BPS vortex solutions governed by the Jackiw-Pi equation are stable in the non-relativistic Chern-Simons matter theory.}

Hereafter we unify expressions of the scalar field $\Psi$ with $\phi$ and refer to it as the Higgs field, and consider the Lagrangian \eqref{eq:Chern-Simons-Lagrangian} governs dynamics of the system.

\subsection{Vortex solutions on torus}\label{sec:basic-JPvortex}

In this section, we quickly review solutions to the Jackiw-Pi equation.
As shown above, the Jackiw-Pi equation is equivalent to Liouville's equation so that arbitrary \revise{meromorphic} functions give local solutions.
For the solutions to be regular vortices, we should pay attention to global behaviour of the field configurations: finiteness of the flux, or the energy, should be imposed.
This requirement certainly gives restrictions for the \revise{meromorphic function in \eqref{eq:solution-JP}.}

Firstly, we consider vortex solutions defined on a flat plane $\mathbb{R}^2$.
It has been shown that the Higgs field $\phi$ given in \eqref{eq:solution-JP} yields the Jackiw-Pi vortices on $\mathbb{R}^2$ if, and only if, the meromorphic function $f$ takes the form
 \begin{align}
     f(z) = \frac{P(z)}{Q(z)},
 \end{align}
where $P(z)$ and $Q(z)$ are polynomials \revise{without common roots} such that $\deg{P}<\deg{Q}$ \cite{Horvathy:1998pe}.
In this case, a vortex number $N$ of the solution is proved to be \revise{twice the degree of $Q$.}

Next, we focus on a case of vortices on a torus $T^2$.
In this case, it is sufficient to impose a doubly periodicity for \revise{the meromorphic function $f$, then the Higgs field also becomes doubly periodic.}
It is a well-known fact that general doubly periodic functions are given in terms of the \revise{Weierstrass} elliptic function $\wp(z)$ and its derivative $\wp'(z)$ on a periodic lattice $\Lambda = 2\mathbb{Z}\omega_1 + 2\mathbb{Z}\omega_2$, where $\omega_1$ and $\omega_2$ are independent complex numbers with positive imaginary part, \revise{called} half-periods.
\revise{Meromorphic functions on this lattice, in general, take the form}
\begin{align}\label{doubly periodic f}
    f(z) = R_1(\wp(z)) + \wp'(z)R_2(\wp(z)),
\end{align}
where $R_1$ and $R_2$ are some rational functions\revise{, see e.g.,} \cite{whittaker_watson_1996}.
Although the general form of meromorphic functions $f(z)$, or Higgs fields, is given, characteristic quantities of vortices such as a vortex number are not obvious from \eqref{doubly periodic f}.
In the next section, we provide analytic calculations of a vortex number on a torus and apply it to specific solutions.
In this approach, we discover detailed structures of the Jackiw-Pi vortices.

We point out here that the Higgs field $\phi(z,\bar{z})$ itself would not be an observable quantity of the non-relativistic Chern-Simons theory. 
However, the flux density $\rho=\abs{\phi}^2$ would be observable.
If this is the case\revise{,} the Higgs field itself is not necessary to be doubly periodic: it only needs quasi-doubly periodicity concerning a lattice $\Lambda = 2\mathbb{Z}\omega_1 + 2\mathbb{Z}\omega_2$, i.e.,
\begin{align}
    \phi(z+2\omega_i) = e^{i\theta_i}\phi(z), \;(i=1,2),
\end{align}
where $\theta_i\in\mathbb{R}$ are some phase angles.
In this context, Akerblom \textit{et.al.} \cite{Akerblom:2009ev} found general doubly periodic solutions for the flux density $\rho$. 
From a gauge theoretical perspective, the quasi-periodic field configurations are admittable because the fields are periodic up to gauge transformation in those cases.
However, we consider in this paper the Jackiw-Pi vortices on $T^2$ constructed from strict doubly periodic Higgs fields $\phi(z,\bar{z})$ itself.
The reason is that a continuity of the Higgs field as a complex function is necessary for the analytic calculation of a vortex number.
On the other hand, the quasi-periodic Higgs fields would give rise to solitonic objects with non-trivial holonomy.
For that kind of vortices, their vortex numbers would be able to take fractional values just as in the case of fractional instantons \cite{tHooft:1981nnx,vanBaal:1982ag,Gonzalez-Arroyo:2019wpu}. 
Although these are interesting solitonic objects, we concentrate here on the trivial holonomy vortices based on the Higgs fields with strict doubly periodicity.

\section{Analytic calculation of vortex number on torus}

In this section, we \revise{perform} the analytic calculation for \revise{a} vortex number of the Jackiw-Pi vortices on a torus.
Then we \revise{consider} several examples including the vortices with simple zero and also multiple zero.
\revise{A similar treatment of the vortex number on a compact manifold is discussed in \cite{Baptista:2012tx} regarding the Taubes vortex. 
We provide details of the approach here.}
An outline of the procedure has been reported in \cite{Miyamoto:2023qwe}\footnote{The report contains inaccuracies based on confusing the ramification points and the poles that contribute to  Higgs zeroes.} by the present authors.

\subsection{Strategy for analysis}

In this section, we provide direct calculations of a vortex number of the Jackiw-Pi vortices on a torus, and observe several new details about the vortex number in terms of a meromorphic function $f(z)$.

The following integral provides a vortex number on a torus.
\begin{equation}\label{eq:integration-F-on-Ttilde}
    N=\frac{1}{2\pi}\int_{\widetilde{T}^2} F_{z\bar{z}} dz\wedge d\bar{z} = \frac{1}{4\pi i} \sum_{i=1}^{m} \oint_{C_{\eta_i}} \pd\log\abs{\phi}^2 dz - \overline{\pd}\log\abs{\phi}^2 d\bar{z},
\end{equation}
where $\widetilde{T}^2$ is a torus without logarithmic singular points $\eta_i$ of the Jackiw-Pi equation, 
\begin{equation}
    \widetilde{T}^2 := T^2\backslash \{\eta_i \in T^2|\phi(\eta_i,\overline{\eta_i})=0\}, \quad (i=1,\dots m).
\end{equation}
The $\eta_i$'s correspond to the zeros of the Higgs field $\phi$, and $C_{\eta_i}$ is a circular boundary around $\eta_i$.
We mention that similar formulations for vortex number in several gauge theories can be found in \cite{Baptista:2012tx, Eto_2006, Gudnason:2021bkw}.

\revise{Let us} evaluate the contour integral \eqref{eq:integration-F-on-Ttilde} around zeroes of $\phi$.
\revise{Note that} from \eqref{eq:solution-JP} the zeroes of $\phi$ may arise from zeroes of $f'(z)$, \revise{namely, ramification points of $f(z)$,} or poles of $|f(z)|^2$.
Hence, we separate the evaluation into two parts: the Higgs zero emerging from the zeroes of $f'(z)$ and from the poles of $|f(z)|^2$.

\revise{Firstly, we} consider the Higgs zeroes emerging from the zeroes of $f'(z)$ with order \revise{$n \geq 1$.}
In such cases, the meromorphic function has the expansion around \revise{$\eta$, the order $n$ zero of $f'(z)$,}
\begin{align}
f(z)=c_0+c_{n+1}(z-\eta)^{n+1}+O\left((z-\eta)^{n+2}\right),\\ f'(z)=(n+1)c_{n+1}(z-\eta)^n+O\left((z-\eta)^{n+1}\right).
\end{align}
\revise{Hence,} the Higgs field is expanded around $\eta$ as
\begin{align}\label{eq:multiple zero}
    \phi(z,\bar{z})=c(z-\eta)^n+O(n+1),
\end{align}
where $c$ is a complex constant composed of the expansion coefficients of $f(z)$\revise{,} and $O(n+1)$ \revise{denotes} the terms of order $n+1$ and higher in $z-\eta, \, \overline{z-\eta}$, and their products.
\revise{On the small circle around $\eta$, we take the parametrisations $z=\eta + \epsilon e^{i\theta},\;\bar{z}=\bar{\eta}+\epsilon e^{-i\theta}$, and $dz=i\epsilon e^{i\theta}d\theta$, where $\epsilon\ll 1$.}
Thus, we find
\begin{align}
   \pd\log\abs{\phi}^2 dz &= \qty(\frac{\pd\phi}{\phi} + \frac{\pd\overline{\phi}}{\overline{\phi}}) dz\nonumber\\
    &\simeq \qty(
        \frac{nc(z-\eta)^{n-1} + O(n)}{c(z-\eta)^n + O(n+1)}+\frac{ O(n)}{\bar{c}(\overline{z-\eta})^n + O(n+1)}  ) dz\nonumber\\
    &\simeq \qty(
        \frac{n}{z-\eta}+ O(0) +\frac{O(n)}{\;(\overline{z-\eta})^n\;}   ) dz
        \xrightarrow[\epsilon\to0]{}ind\theta,
\end{align}
and 
\begin{align}
        \overline{\pd}\log\abs{\phi}^2 d\bar{z} &=\qty(\frac{\overline{\pd}\phi}{\phi} + \frac{\overline{\pd}\, \overline{\phi}}{\overline{\phi}}) d\bar{z}\nonumber\\
    &\simeq  \qty(
        \frac{O(n)}{c (z-\eta)^n+O(n+1)} + \frac{n\overline{c}(\overline{z-\eta})^{n-1} + O(n)}{\overline{c}(\overline{z-\eta})^n + O(n+1)}) d\bar{z}\nonumber\\
    &\simeq  \qty(
        \frac{O(n)}{(z-\eta)^n} + \frac{n}{\;(\overline{z-\eta})^n\;} + O(0)) d\bar{z} \xrightarrow[\epsilon\to 0]{} -ind\theta.
\end{align}
From this expansion, the contribution to the vortex number from order $n$ zeroes is
\begin{align}\label{eq:vortex_number_from_n_zeros}
    N(\mbox{order $n$ zero})&=\frac{1}{4\pi i}\sum_{\eta_i}\oint_{C_{\eta_i}}
     \pd\log\abs{\phi}^2 dz- \overline{\pd}\log\abs{\phi}^2 d\bar{z}\nonumber\\
     &\xrightarrow[\epsilon\to0]{} \frac{1}{4\pi i}\sum_{\eta_i}\oint_{C_{\eta_i}}2ind\theta=n\times(\mbox{Number of order $n$ zeroes}).
\end{align}
Hence the Higgs zero of order $n$ emerged from an $n$-th zero of $f'$ contributes to the vortex number $n$.

\revise{Secondly,} we consider the Higgs zeroes that emerged from the poles of the meromorphic function $f(z)$.
Assuming that $f(z)$ has an order $n$ pole at $z=\eta$ with $n\geq 1$, the Laurent expansion of $f(z)$ around $\eta$ is
\begin{align}
    f(z)=\frac{c}{(z-\eta)^n}+O\left((z-\eta)^{-n+1}\right),
\end{align}
then its derivative is
\begin{align}
    f'(z)=-\frac{nc}{(z-\eta)^{n+1}}+O\left((z-\eta)^{-n}\right),
\end{align}
where $c$ is a complex coefficient and $O\left((z-\eta)^{-n}\right)$ \revise{denotes} the terms of order $-n$ and higher in $z-\eta$.
Note that the singularities of $f(z)$ are only poles since it is meromorphic. 
To find the behaviour of the Higgs zeroes, we observe that
\begin{align}
    1+|f(z)|^2&=1+\left(\frac{c}{(z-\eta)^n}+\widetilde{O}(-n+1)\right)\left(\frac{\overline{c}}{(\overline{z-\eta})^n}+\widetilde{O}(-n+1)\right)\nonumber\\
    &=1+\frac{|c|^2}{(z-\eta)^n(\overline{z-\eta})^n}\left(1+\widetilde{O}(1)\right)
    =\frac{|c|^2}{(z-\eta)^n(\overline{z-\eta})^n}\left(1+\widetilde{O}(1)\right),
\end{align}
where $\widetilde{O}(n)$ is the terms of order $n\in\mathbb{Z}$ and higher in $z-\eta,\;\overline{z-\eta}$, and their products.
Note that \revise{$\widetilde{O}(n)$} may include negative powers of $z-\eta$ and $\overline{z-\eta}$ such as $(\overline{z-\eta})^{n+m}/(z-\eta)^m$ in contrast to \revise{$O(n)$ in the former case.}

Thus the behaviour of the Higgs field $\phi$ around $\eta$ is
\begin{align}\label{eq:Higgs zero from poles}
   \phi(z,\bar{z})&= \frac{f'(z)}{1+|f(z)|^2}\simeq\left(-\frac{nc}{(z-\eta)^{n+1}}+\widetilde{O}(-n)\right)\frac{(z-\eta)^n(\overline{z-\eta})^n}{|c|^2}\left(1+\widetilde{O}(1)\right)\nonumber\\
    &=\left(-\frac{n(\overline{z-\eta})^n}{\bar{c}(z-\eta)}+\widetilde{O}(n)\right)\left(1+\widetilde{O}(1)\right)\nonumber\\
    &=-\frac{n(\overline{z-\eta})^n}{\bar{c}(z-\eta)}+\widetilde{O}(n),
\end{align}
and similarly,
\begin{align}\label{eq:Higgs-bar zero from poles}
    \overline{\phi}(z,\bar{z})
    =-\frac{n(z-\eta)^n}{c(\overline{z-\eta})}+\widetilde{O}(n).
\end{align}
We find from \eqref{eq:Higgs zero from poles} that the zeroes of the Higgs field $\phi$ emerge from the cases $n\geq2$, while the $n=1$ case gives a non-zero point.
The contour integral around $\eta$ for the $n=1$ case does not contribute to the vortex number and the point $z=\eta$ becomes a saddle point of $\phi$ because there remains no radius dependence in its leading order at this point, as we will see later in some examples.
We thus observe that
\begin{align}
    \frac{\partial \phi}{\phi}\simeq \frac{\displaystyle\frac{n(\overline{z-\eta})^n}{\overline{c}(z-\eta)^2}+\widetilde{O}(n-1)}{\displaystyle -\frac{n(\overline{z-\eta})^n}{\bar{c}(z-\eta)}+\widetilde{O}(n)}=-\frac{1}{z-\eta}+\widetilde{O}(0),
\end{align}
and
\begin{align}
    \frac{\partial\overline{\phi}}{\overline{\phi}}\simeq \frac{\displaystyle -\frac{n^2(z-\eta)^{n-1}}{c(\overline{z-\eta})}+\widetilde{O}(n-1)}{\displaystyle -\frac{n(z-\eta)^n}{c(\overline{z-\eta})}+\widetilde{O}(n)}=\frac{n}{z-\eta}+\widetilde{O}(0),
\end{align}
around $\eta$, respectively.
The $z$-derivative part of the contour integral of \eqref{eq:integration-F-on-Ttilde} and its $\overline{z}$-derivative counterpart for these cases become
\begin{align}
    \oint_{C_\eta}\pd\log\abs{\phi}^2 dz &=\oint_{C_\eta} \qty(\frac{\pd\phi}{\phi} + \frac{\pd\overline{\phi}}{\overline{\phi}}) dz\nonumber\\
    &=\oint_{C_\eta} \left(\frac{(n-1)}{z-\eta}+\widetilde{O}(0)\right)dz\xrightarrow[\epsilon\to0]{}2(n-1)\pi i,
\end{align}
and 
\begin{align}
    \oint_{C_\eta}\overline{\pd}\log\abs{\phi}^2 d\overline{z} &=\oint_{C_\eta} \qty(\frac{\overline{\pd}\phi}{\phi} + \frac{\overline{\pd}\overline{\phi}}{\overline{\phi}}) d\overline{z}\xrightarrow[\epsilon\to0]{}-2(n-1)\pi i,
\end{align}
respectively, where the parametrization $z=\eta+\epsilon e^{i\theta},$ etc., is applied as in the former \revise{case.} 
Note that the $\widetilde{O}(0)$ term such as $(\overline{z-\eta})dz/(z-\eta)$ does not contribute to the contour integral around $\eta$ because it has order $\epsilon$.
Hence the contour integral is evaluated as
\begin{align}\label{eq:vortex number from n-th pole}
 N(\mbox{order $n$ pole})&=\frac{1}{4\pi i}\sum_{\eta_i}\oint_{C_{\eta_i}}
     \pd\log\abs{\phi}^2 dz- \overline{\pd}\log\abs{\phi}^2 d\bar{z}\nonumber\\
     &\xrightarrow[\epsilon\to0]{} \frac{1}{4\pi i}\sum_{\eta_i}\oint_{C_{\eta_i}}2i(n-1)d\theta=(n-1)\times(\mbox{Number of order $n$ poles}).
\end{align}
Therefore, we find that the Higgs zero emerged from the order $n$ poles of $f$ contributes to the vortex number $n-1$.
We note \revise{again} that the order $1$ poles of $f$ do not contribute to the vortex number since they do not give the zero points of $\phi$, namely, not the vortex centres.

To summarize this subsection, we have \revise{provided} \revise{the} analytical \revise{calculations of} the vortex number of the Jackiw-Pi vortex on a torus.
In this approach, the defining region of the flux is regarded as a torus without the singular points corresponding to the centres of the vortex, namely, the Higgs zeroes.
The evaluation of the flux integration is given through the expansion around each singular point and contour integration.
It is shown that there are two types of Higgs zeroes: One of them arises from the zeroes of the derivative of a meromorphic function $f'$, and the other emerges from the poles of $f$.
The contribution to the vortex number is $n$ from the order $n$ zeroes of $f'$, and $n-1$ from the order $n$ poles of $f$.

\subsection{Vortices from elliptic functions: Examples}

\revise{In this section, we} examine typical examples of the Jackiw-Pi vortices on a torus.
We choose simple elliptic functions as a meromorphic function $f(z)$ and see the characteristic aspects of the Higgs fields for individual vortices.
In particular, we perform analytic calculations for the vortex number \revise{studied} in the previous subsection and see the consistency with the numerical integration.

It is well known that there are two kinds of fundamental doubly periodic functions, namely, the Weierstrass $\wp$ function and the Jacobi elliptic function.
We will find that the characteristic difference appears between the vortices constructed from these fundamental elliptic functions.

\paragraph{Example 1: Weierstrass $\wp$ function}
First of all, let us consider the Weierstrass $\wp$ function  $\wp(z;\omega_1,\omega_2)$ for the meromorphic function $f$, then the Higgs field becomes
\begin{equation}\label{eq:Higgs from pe}
    \phi_{\wp}(z,\bar{z}) = \frac{\wp'(z;\omega_1,\omega_2)}{1+\abs{\wp(z;\omega_1,\omega_2)}^2},
\end{equation}
where $\omega_1,\omega_2\in\mathbb{C}$ are the half-periods of \revise{a} lattice\footnote{In some literature these are defined as the whole-periods.} $\Lambda = 2\mathbb{Z}\;\omega_1 + 2\mathbb{Z}\;\omega_2$, and $\wp(z;\omega_1,\omega_2)$ is doubly periodic with respect to the lattice.
The $\wp$-function enjoys the differential equation
\begin{align}\label{eq:differential eq for pe}
    \wp'{\,}^2(z)=4\wp^3(z)-g_2\wp(z)-g_3,
\end{align}
where $g_2$ and $g_3$ are constants and the dependence of the $\omega_1$ and $\omega_2$ is omitted.
The values of $\wp$-function at its half-periods are denoted as $e_j=\wp(\omega_j),\,(j=1,2,3)$ with  $\omega_3:=\omega_1+\omega_2$, and there exist relations $e_1e_2+e_2e_3+e_3e_1=-g_2/4$, and $e_1e_2e_3=g_3/4$.
It is known that the $\wp$ function has a double pole at the origin such as
\begin{align}
    \wp(z)=\frac{1}{z^2}+O(z^0).
\end{align}
Hence the Higgs field has a zero with a unit vortex number at the origin from the discussion of the previous subsection.
We find from \eqref{eq:Higgs zero from poles} that the phase angle rotates three times around the origin, 
\begin{align}
    \phi_\wp (z,\overline{z})&\simeq -\frac{2\overline{z}^2}{z}+\widetilde{O}(2)\nonumber\\
    &=-2\epsilon e^{-3\theta i}+O(\epsilon^2),
\end{align}
where $z=\epsilon e^{i\theta}$ and $\overline{z}=\epsilon e^{-i\theta}$ are employed.

On the other hand, the $\wp'$ has simple zeroes at the half-period points $z=\omega_1,\,\omega_2$ and $\omega_3$, so that there exist three simple zeroes of the Higgs field at that points.
We observe that\revise{, from \eqref{eq:multiple zero},} the phase angle rotates once around these simple zeroes,
\begin{align}
    \phi_\wp (z,\overline{z})\simeq c(z-\omega_j)+O(2)=c\epsilon e^{i\theta}+O(\epsilon^2),
\end{align}
where $c=(6e_j^2-g_2/2)/(1+|e_j|^2)$ with $j=1,2,$ and $3$.
Thus we find these zeroes, namely the vortex centres, contribute to the vortex number $3$ from \eqref{eq:vortex_number_from_n_zeros} so that the total vortex number of this solution is $4$.
This is consistent with the numerical integration \revise{of the magnetic flux} \eqref{eq:first Chern number} performed with Mathematica,
\begin{align}
   \frac{1}{2\pi}\int F_{xy} dxdy=\frac{1}{2\pi}\int 4|\phi_\wp|^2 dxdy =4,
\end{align}
where \eqref{eq:JP_equation_pre} is applied.
\begin{figure}[htbp]
    \centering
    \includegraphics[width=0.55\linewidth]{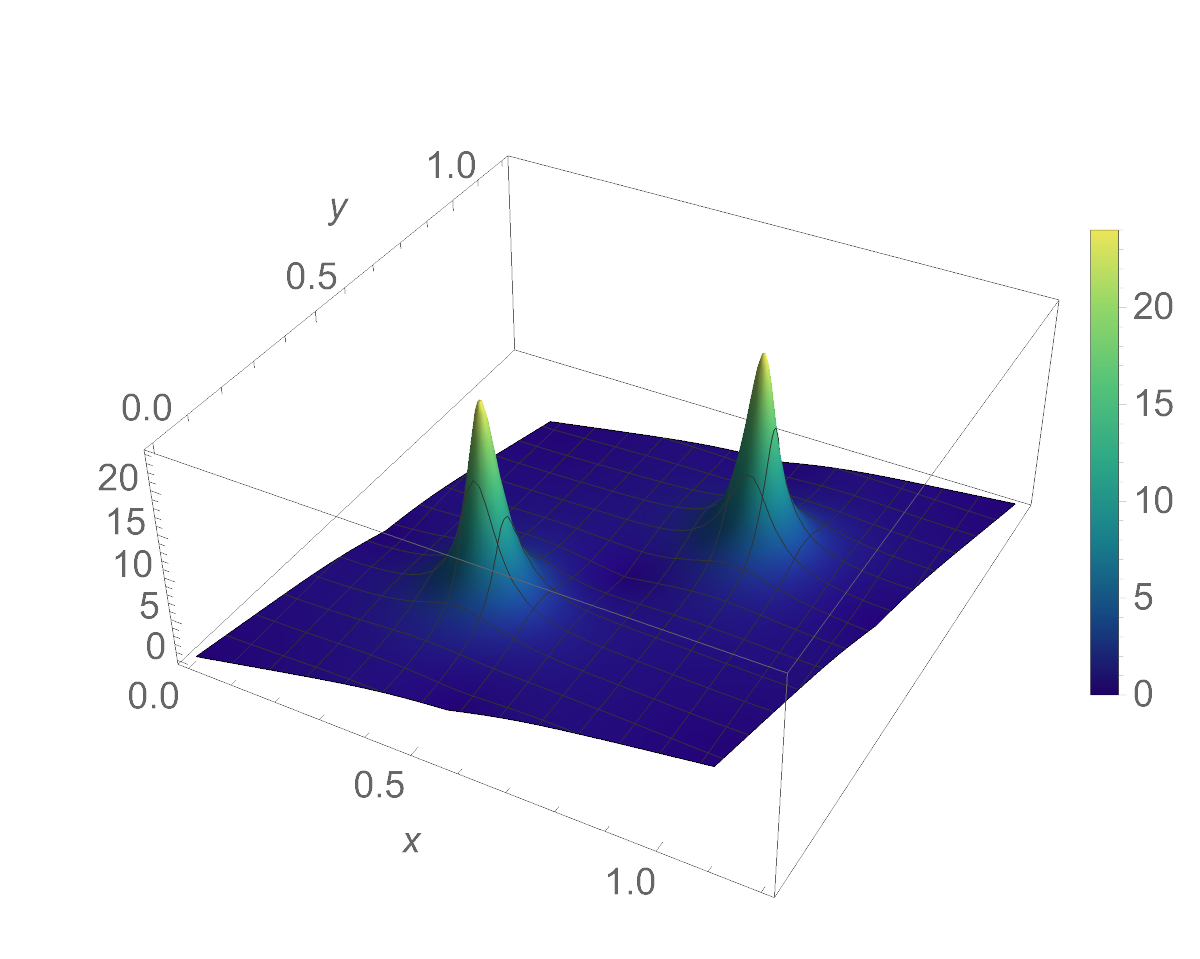}
    \hspace{3mm}
    \includegraphics[width=0.40\linewidth]{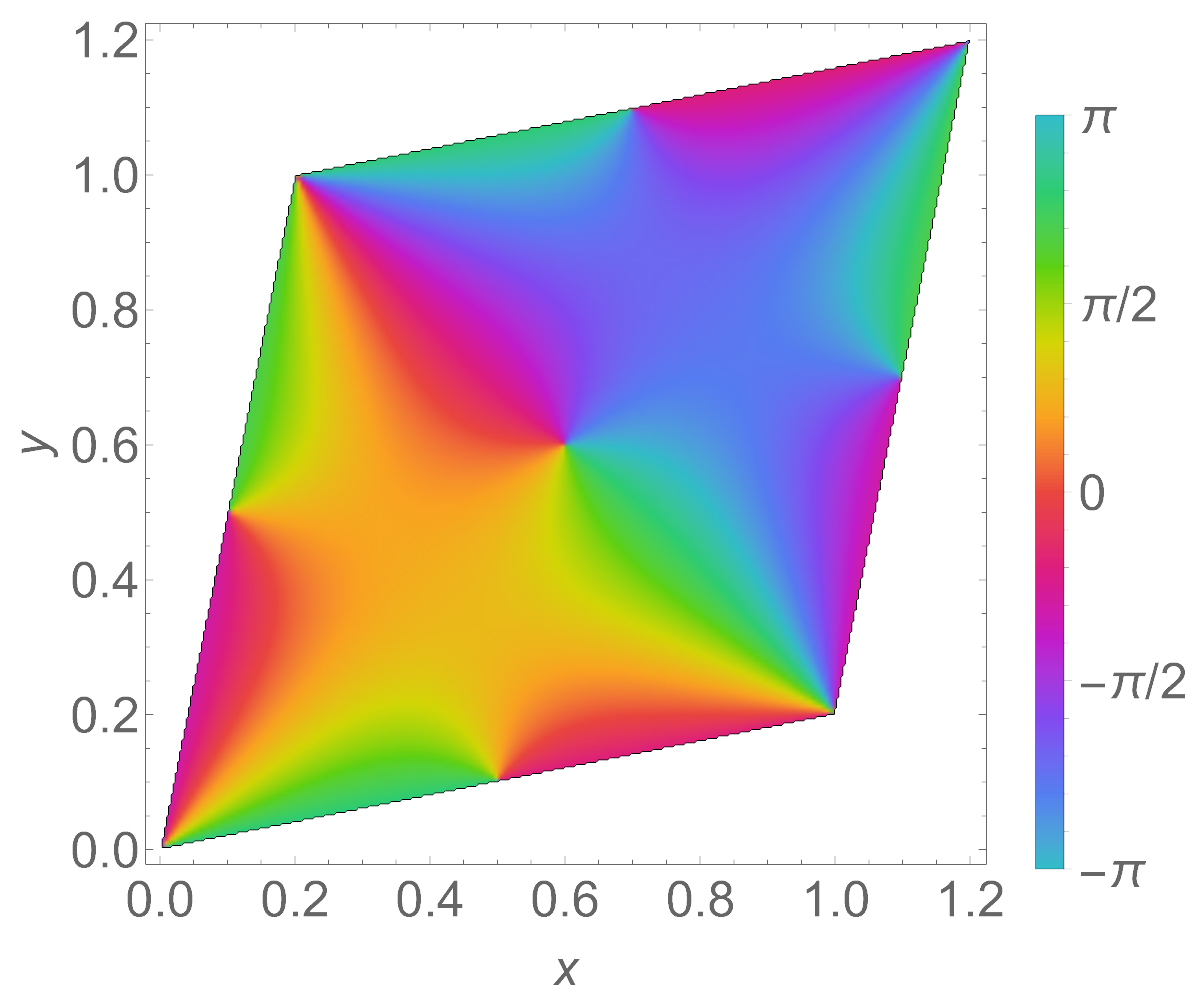}
    \caption{
    \revise{Profile of $\abs{\phi_{\wp}}$ on the fundamental parallelogram (left), and its phase angle(right).}
    }
    \label{fig:phi-wp-3D}
\end{figure}

Figure \ref{fig:phi-wp-3D} shows the profile of the Higgs field $\phi_\wp$ with half-periods $\omega_1=0.5+0.1 i$ and $\omega_2=0.1+0.5i$.
The four \revise{zeroes} exist at $0, \omega_1, \omega_2,$ and $\omega_3$ in the fundamental lattice.
The phase angle structure shows the three times rotation around $z=0$, while the rotations are once around the other zeroes at $z=\omega_i,\,(i=1,2,3)$, as expected. 

\begin{figure}[htbp]
    \centering
    \includegraphics[width=0.4\linewidth]{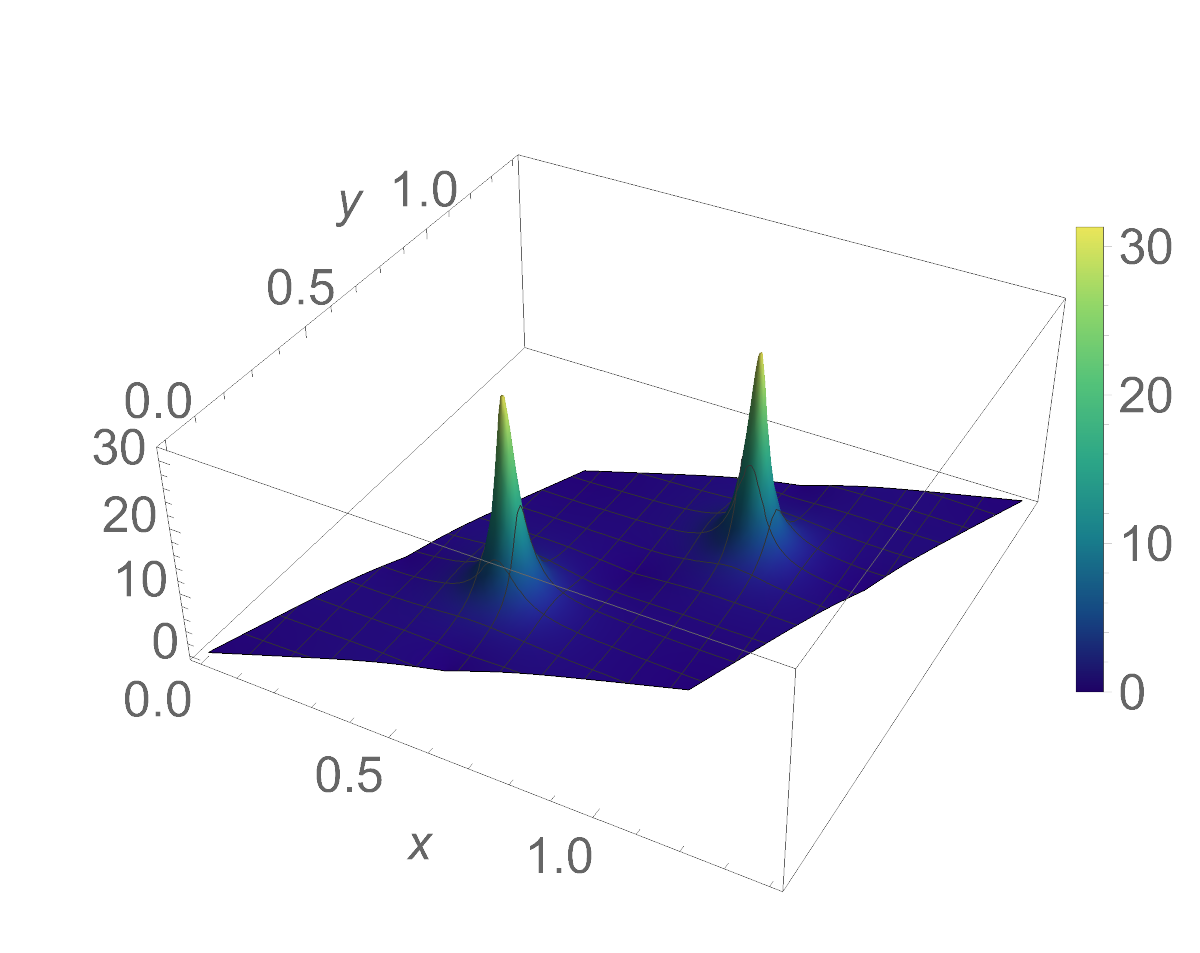}
    \includegraphics[width=0.3\textwidth]{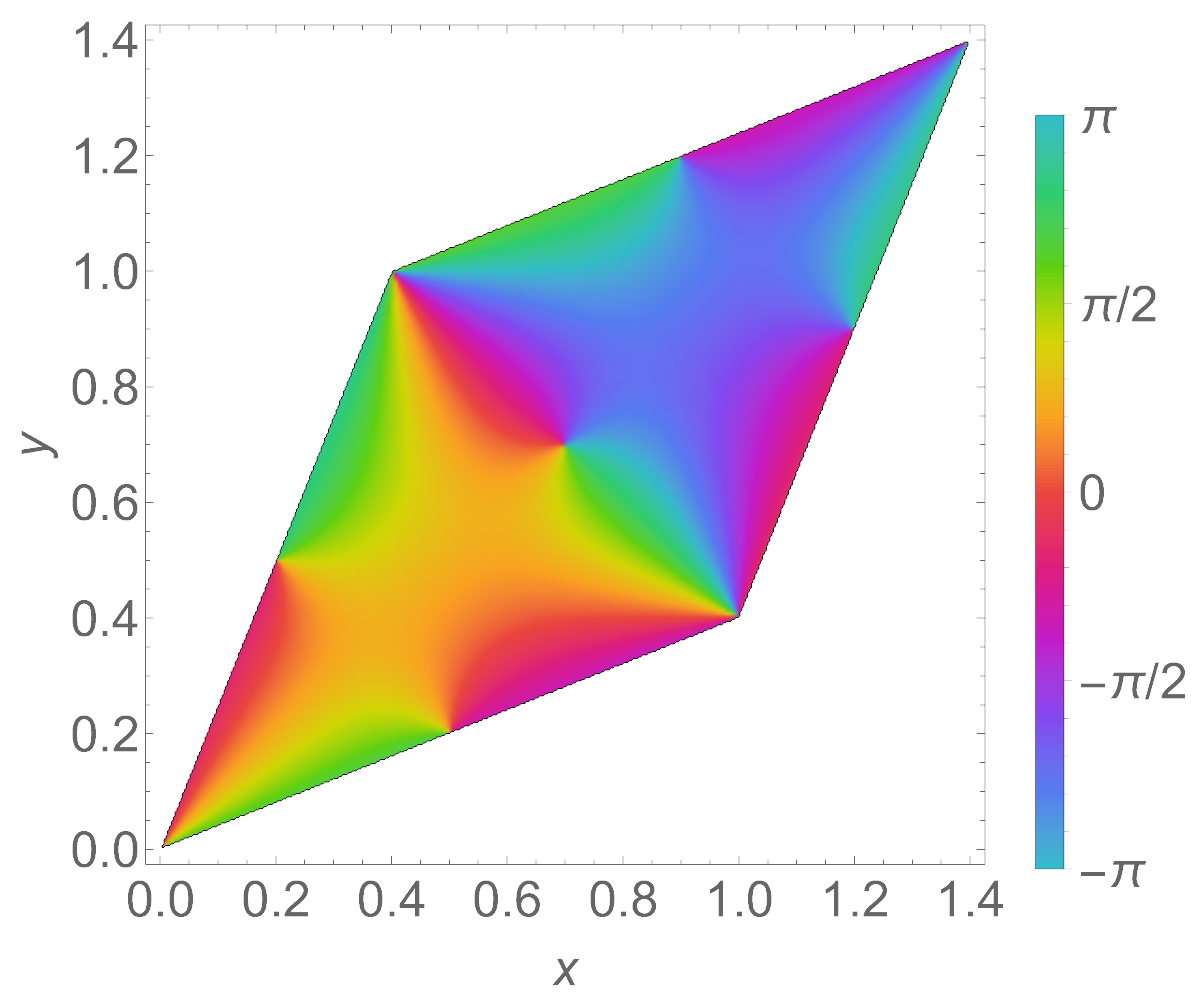}
    \includegraphics[width=0.4\textwidth]{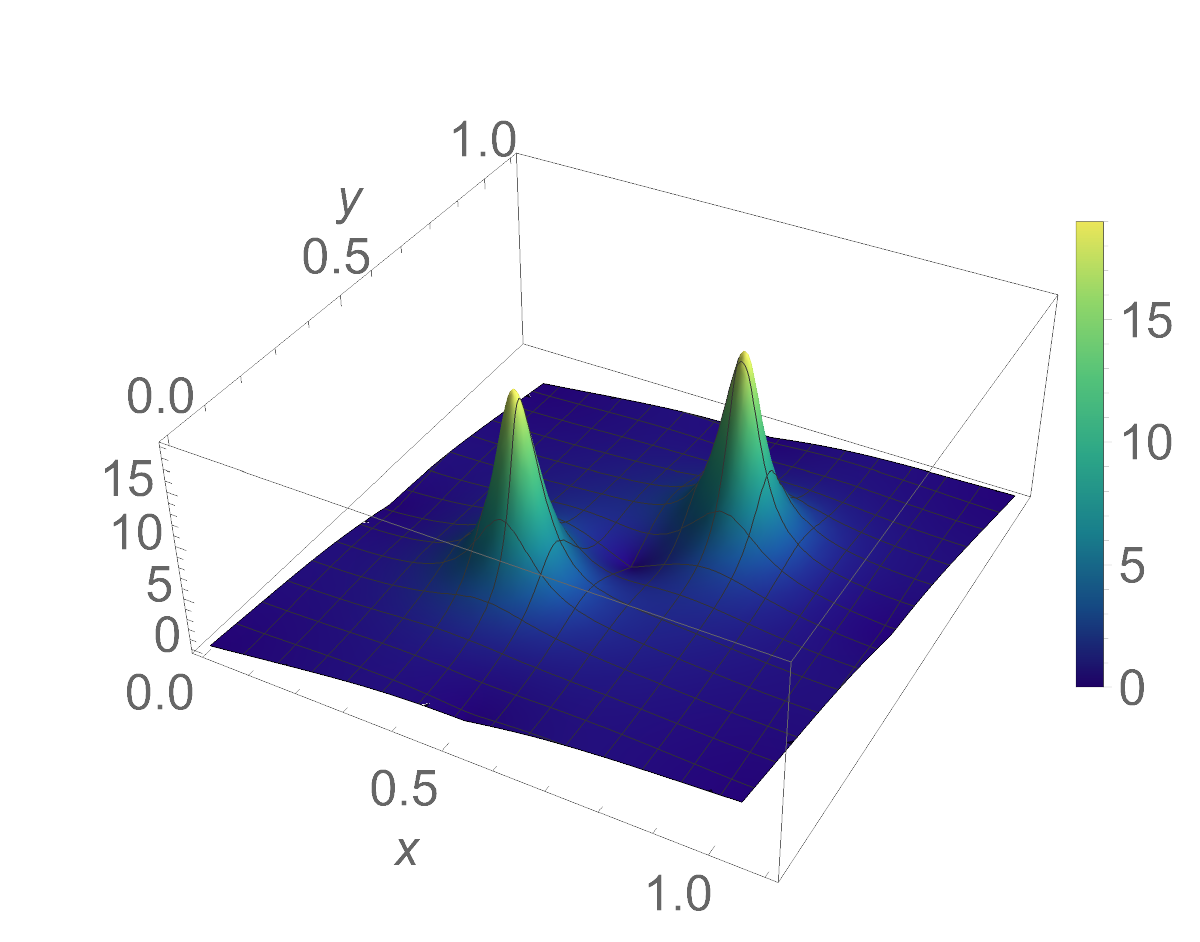}
    \includegraphics[width=0.3\textwidth]{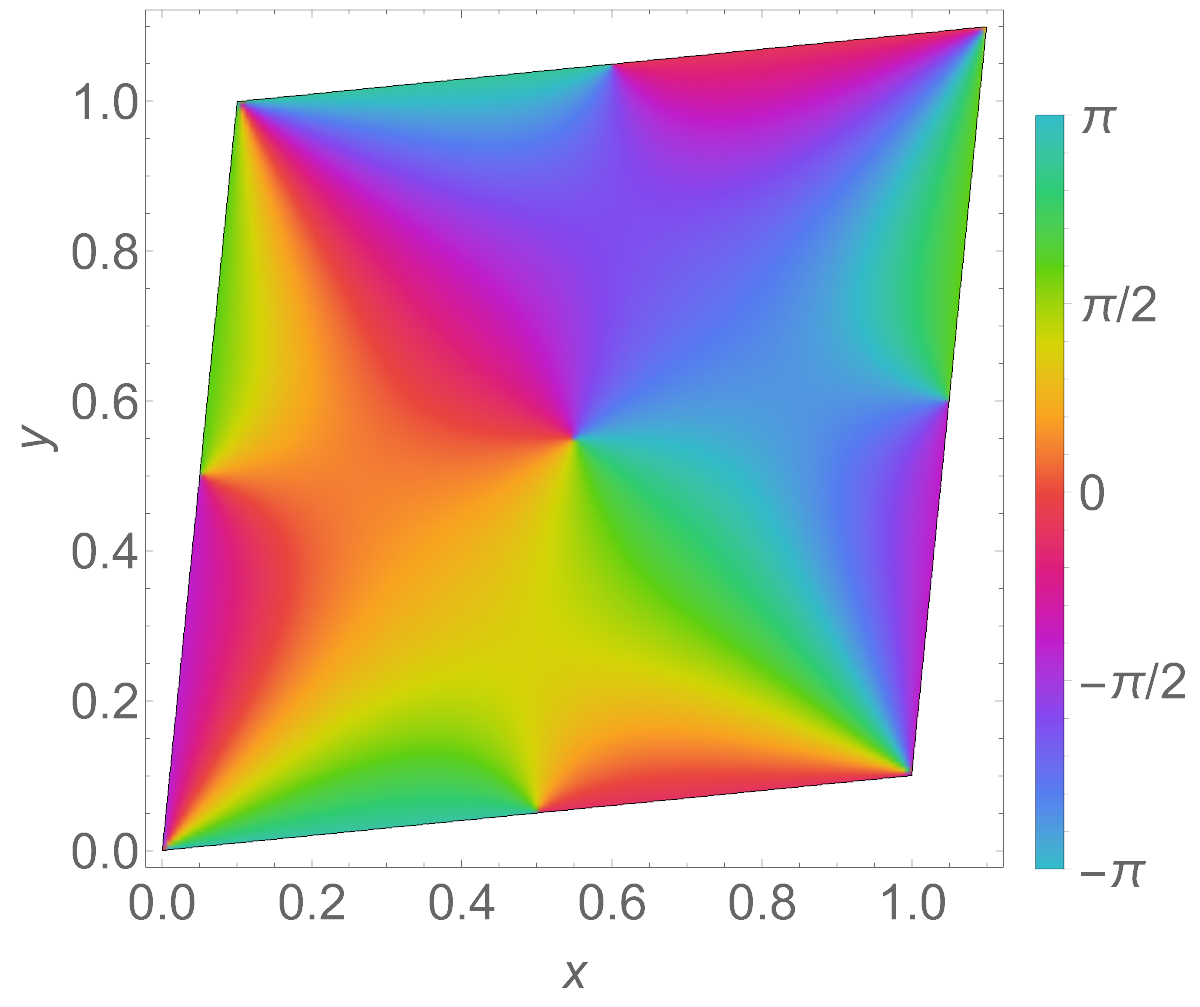}
    \includegraphics[width=0.4\textwidth]{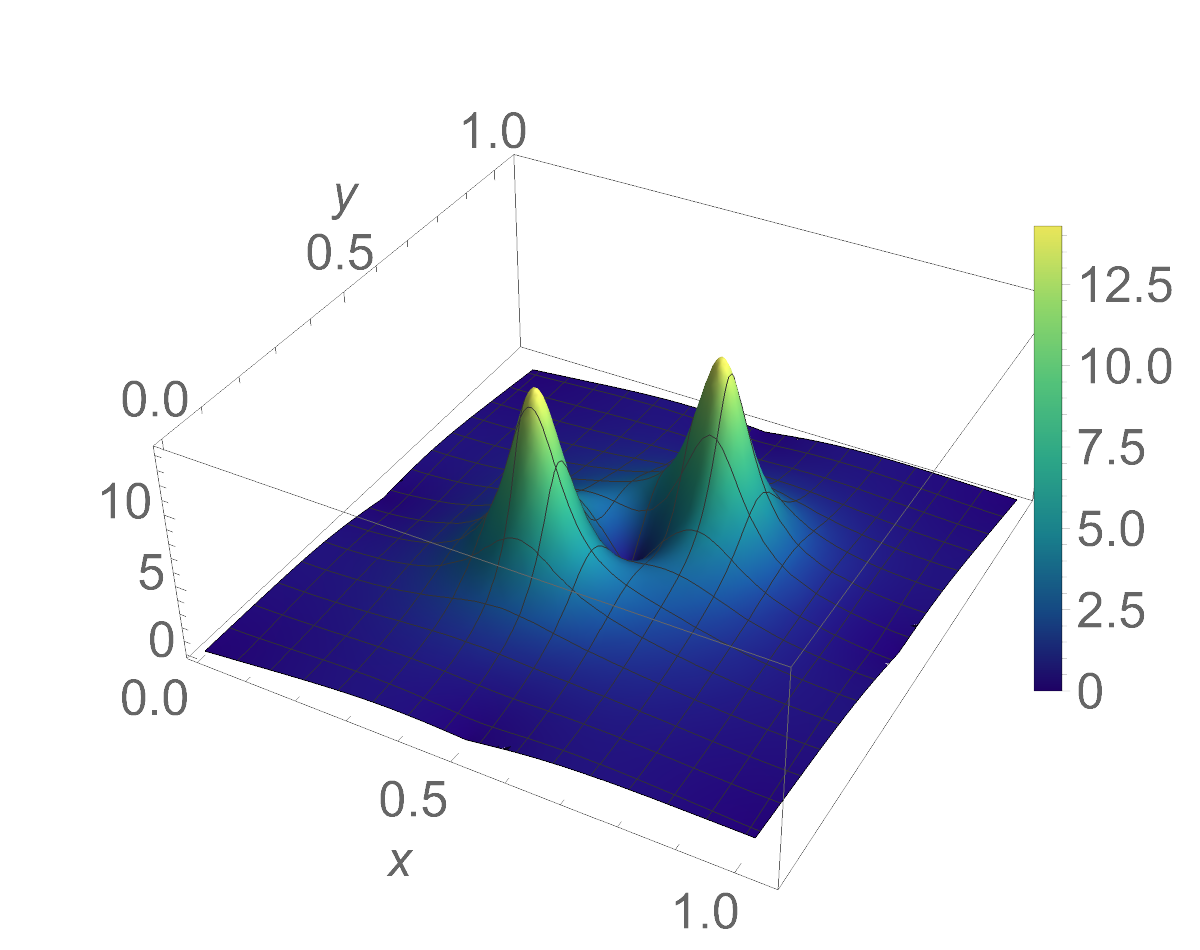}
    \includegraphics[width=0.3\textwidth]{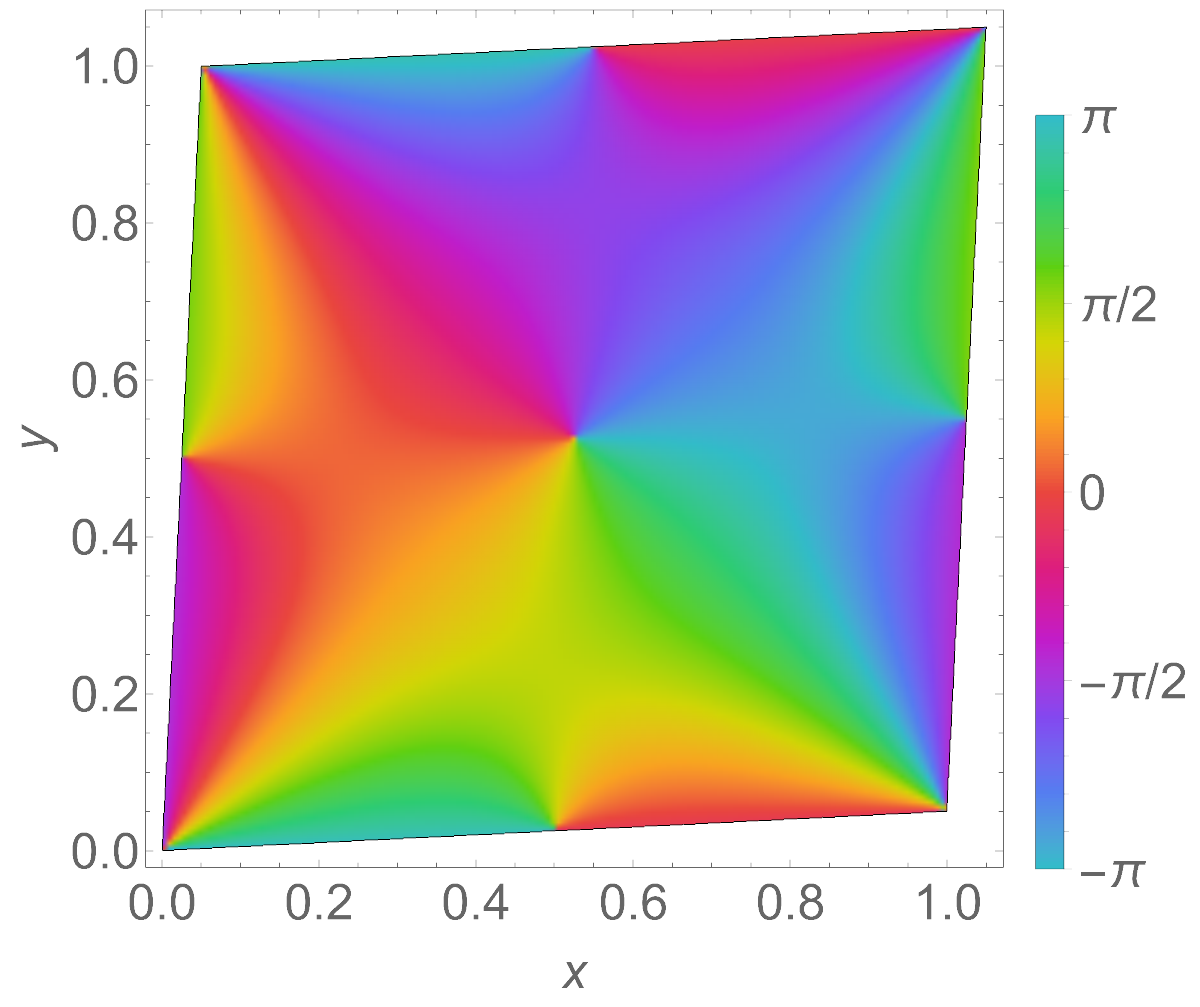}
    \includegraphics[width=0.4\textwidth]{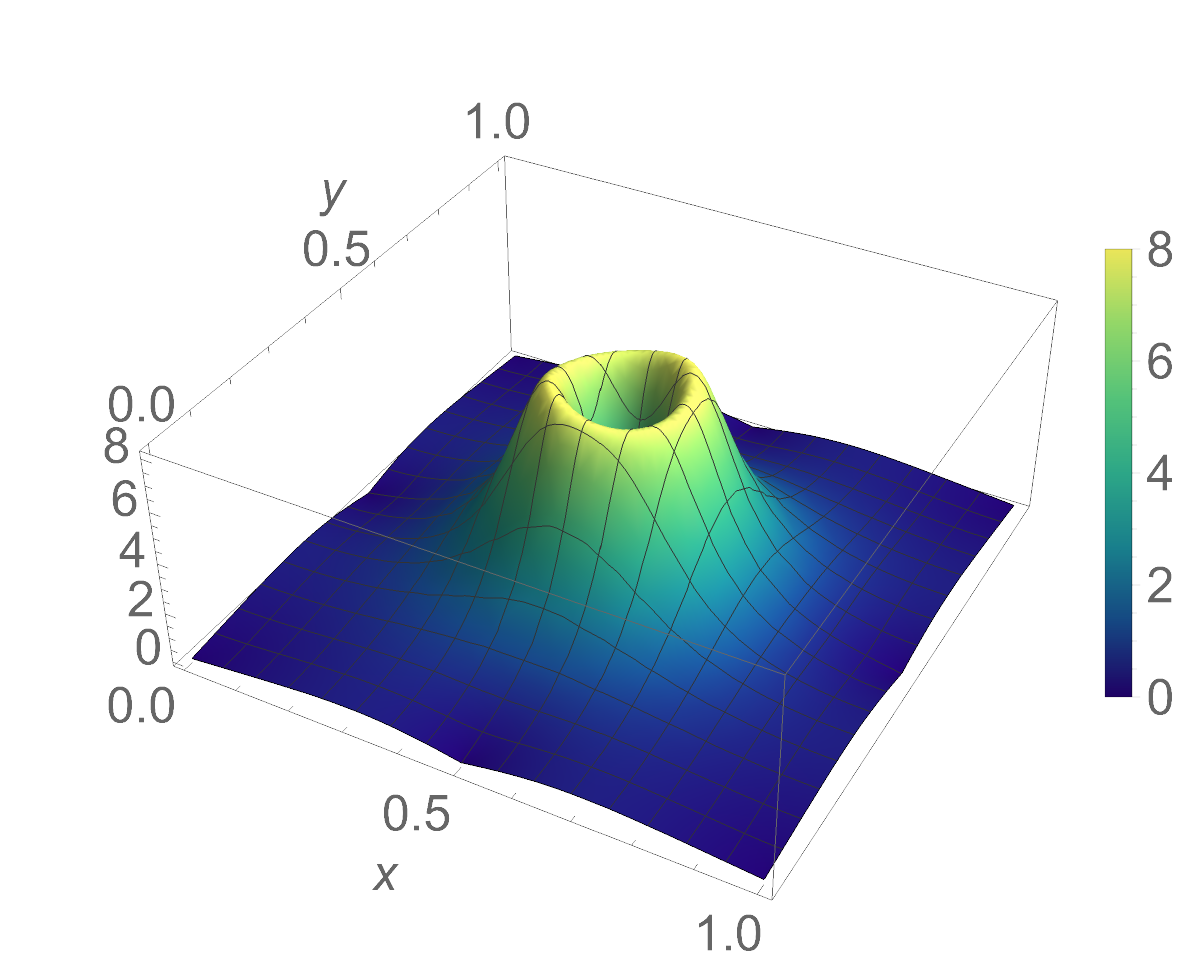}
    \includegraphics[width=0.3\textwidth]{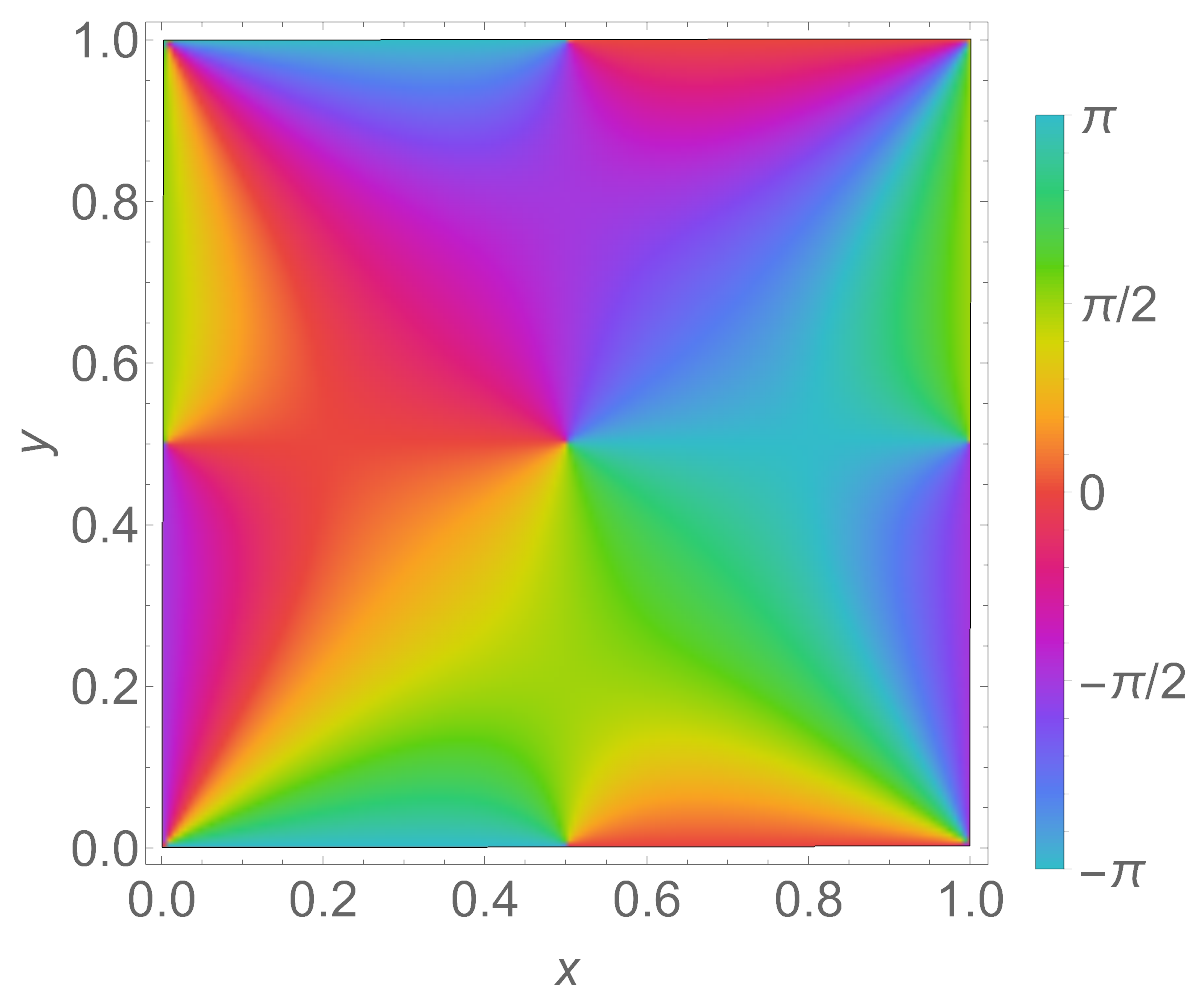}
    \caption{Profiles of $\abs{\phi_{\wp}}$ (left column) and its phase angle (right column) with $(\omega_1,\omega_2)=(0.5+0.2 i, 0.2+0.5 i)$ (top), $(\omega_1,\omega_2)=(0.5+0.05i,0.05+0.5i)$ (second), $(\omega_1,\omega_2)=(0.5+0.025i,0.025+0.5i)$ (third), $(\omega_1,\omega_2)=(0.5+0.001i,0.001+0.5i)$ (bottom).
    The vertical scale is not unified.}
    \label{fig:phi_wp_periods}
\end{figure}

The absolute value of the Higgs field indicates that the flux is localised at the twin peaks on the fundamental lattice.
This localisation structure is not the common feature of $\phi_\wp$ but depends on the choice of half-periods as illustrated in Figure \ref{fig:phi_wp_periods}.
As the region of the fundamental lattice tends to be rectangular, the twin peaks eventually merge into a ``volcano-like" structure surrounding the zero at $\omega_3$.

\paragraph{Example 2: Jacobi sn function}
The other fundamental elliptic function is the Jacobi elliptic function.
Now we apply the Jacobi $\sn$ function simply as the meromorphic function $f$, then the Higgs field takes the form
\begin{equation}\label{eq:higgs from sn}
    \phi_{\sn}(z,\bar{z}) = \frac{\cn(z;k)\dn(z;k)}{1+\abs{\sn(z;k)}^2},
\end{equation}
where $\sn'(z;k)=\cn(z;k)\dn(z;k)$ and $k$ is the modulus.
The fundamental lattice on which the $\sn$ function defined is $\Lambda=4K(k)\mathbb{Z}+2iK'(k)\mathbb{Z}$, where $K(k)=\int_0^{\pi/2} d\theta \qty(1-k^2\sin^2 \theta)^{-1/2}$ and $K'(k):=K(k')$ are the complete elliptic integrals of first kind, with $k'^2=1-k^2$.
We take the modulus $k\in[0,1)$ as usual, for which $K(k),\,K'(k)>0$ so that the fundamental lattice is rectangular.

This vortex solution has four simple zeroes emerging from the simple zeroes of the numerator in \eqref{eq:higgs from sn}, i.e., $K$ and $3K$ are the zeroes of $\cn$, and $3K+iK'$ and $3K+iK'$ are those of $\dn$.
In contrast to the $\wp$-function case, the $\sn$-function in the denominator has only simple poles at $iK'$ and $2K+iK'$ so they do not give zero points but saddle points from \eqref{eq:Higgs zero from poles}.

\begin{figure}[htbp]
    \centering
    \includegraphics[width=0.4\textwidth]{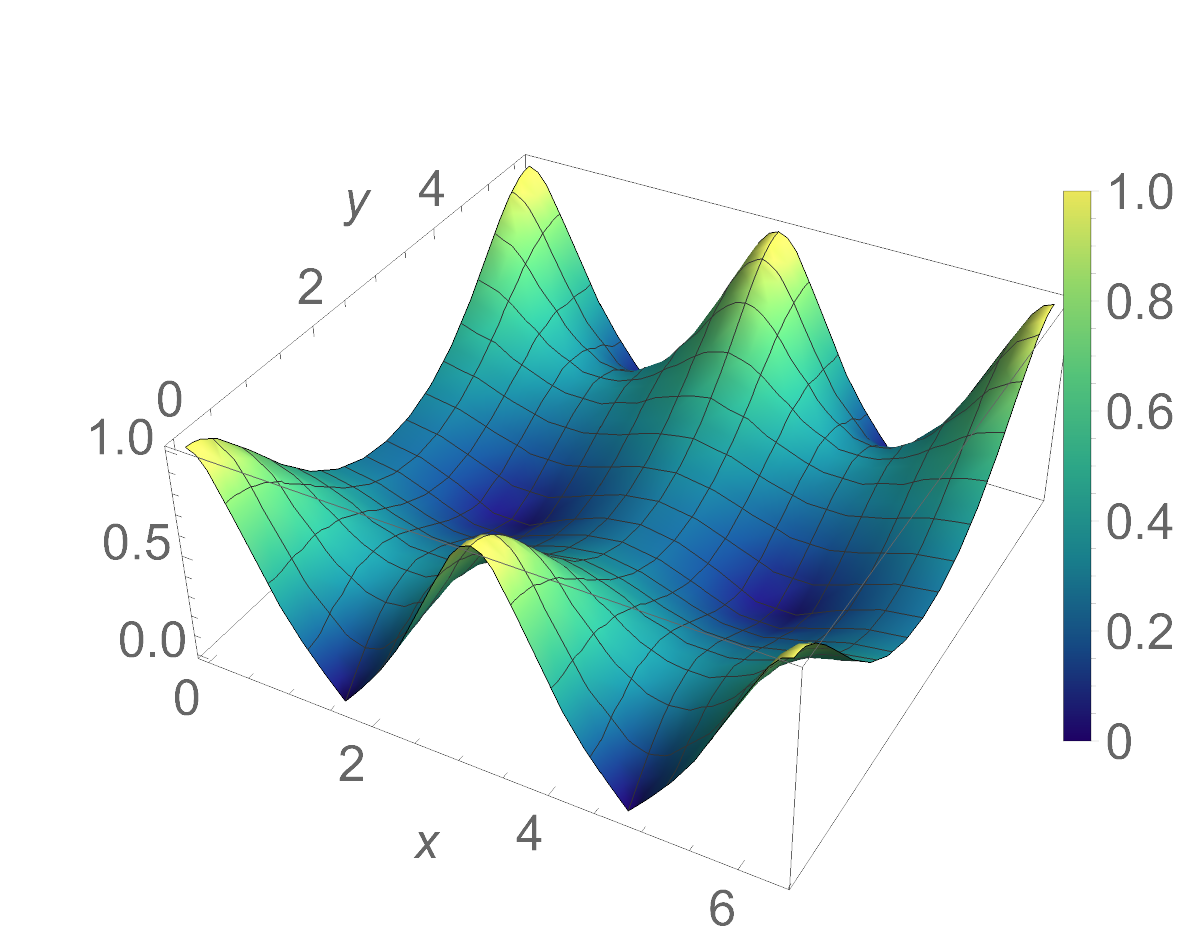}
    \includegraphics[width=0.3\textwidth]{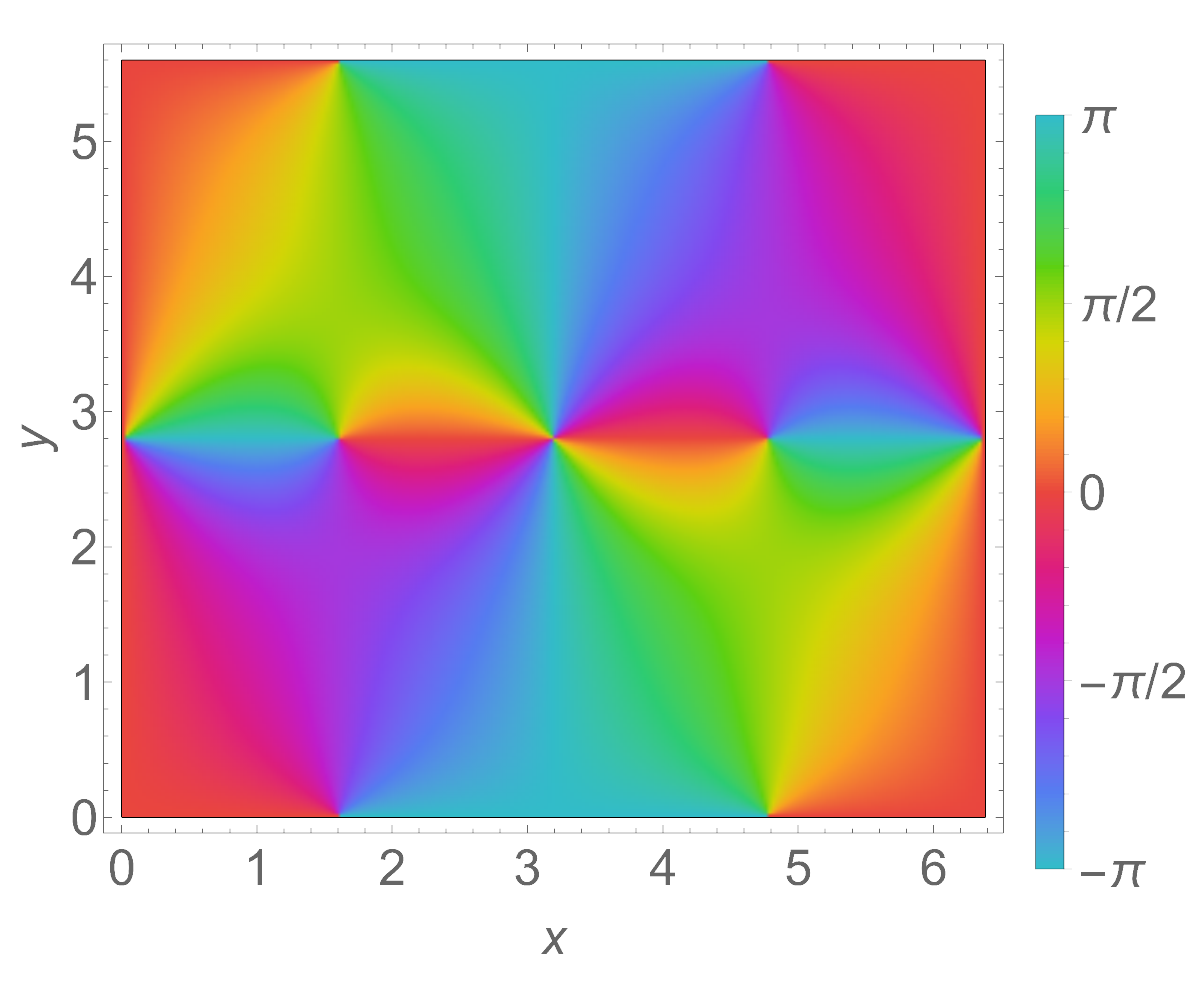}
    \includegraphics[width=0.4\textwidth]{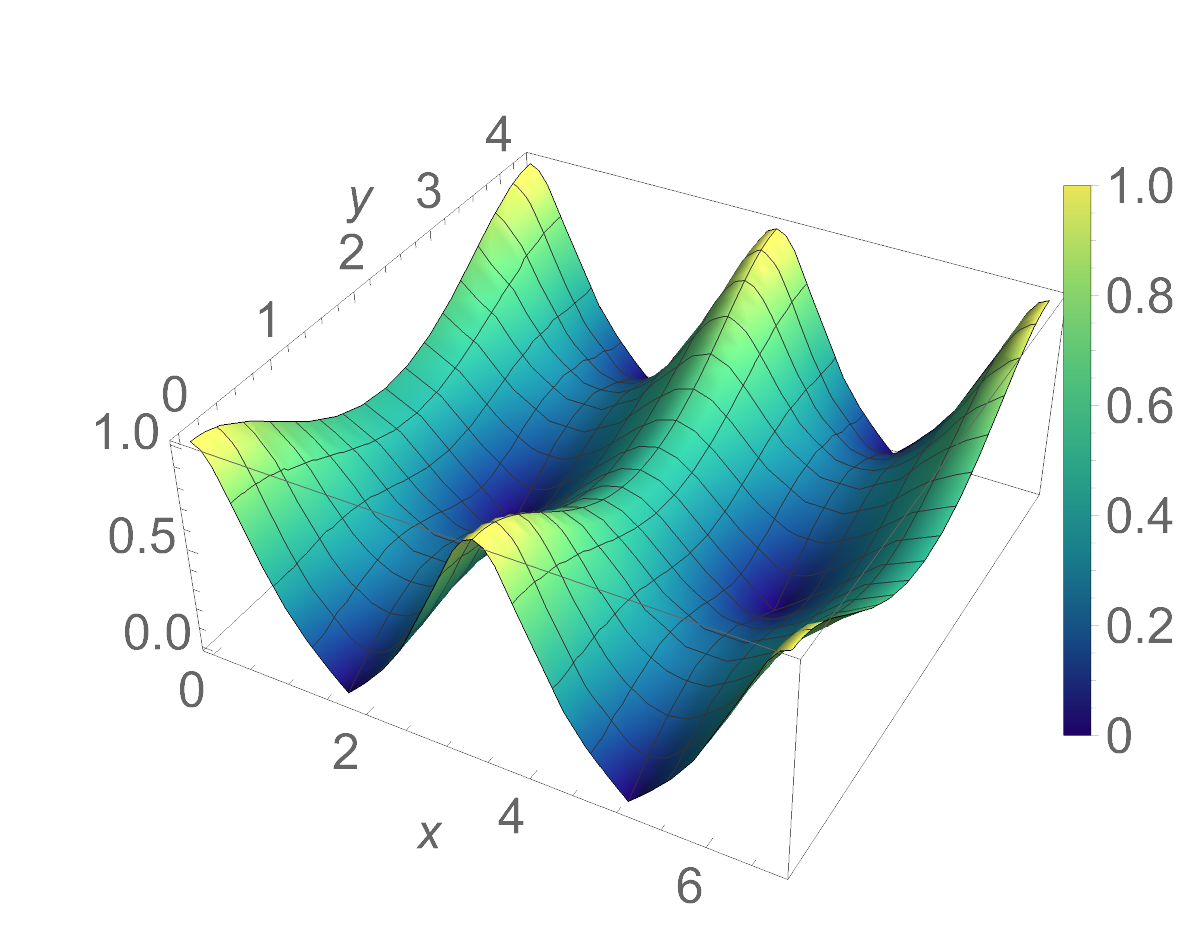}
    \includegraphics[width=0.3\textwidth]{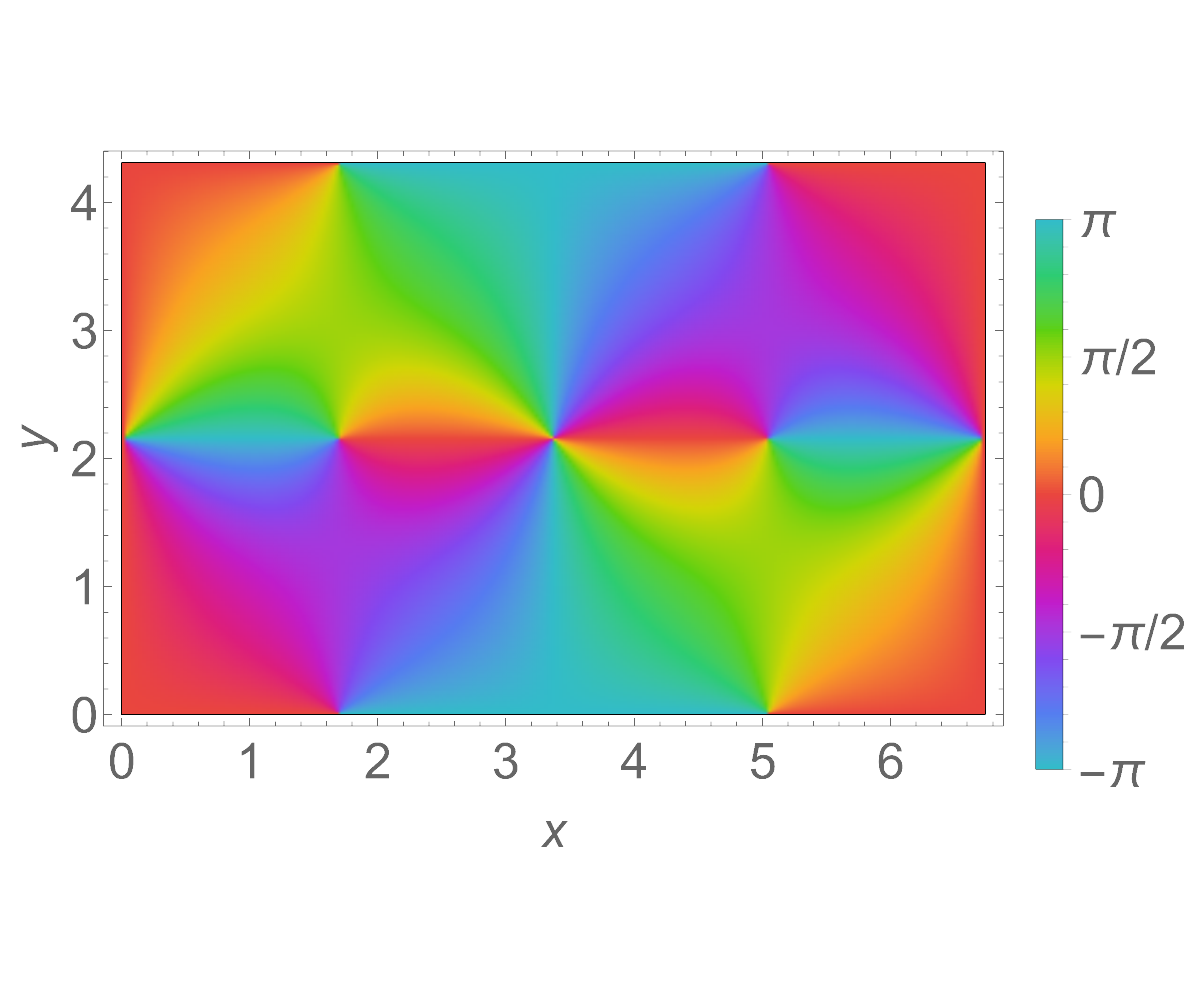}
    \includegraphics[width=0.4\textwidth]{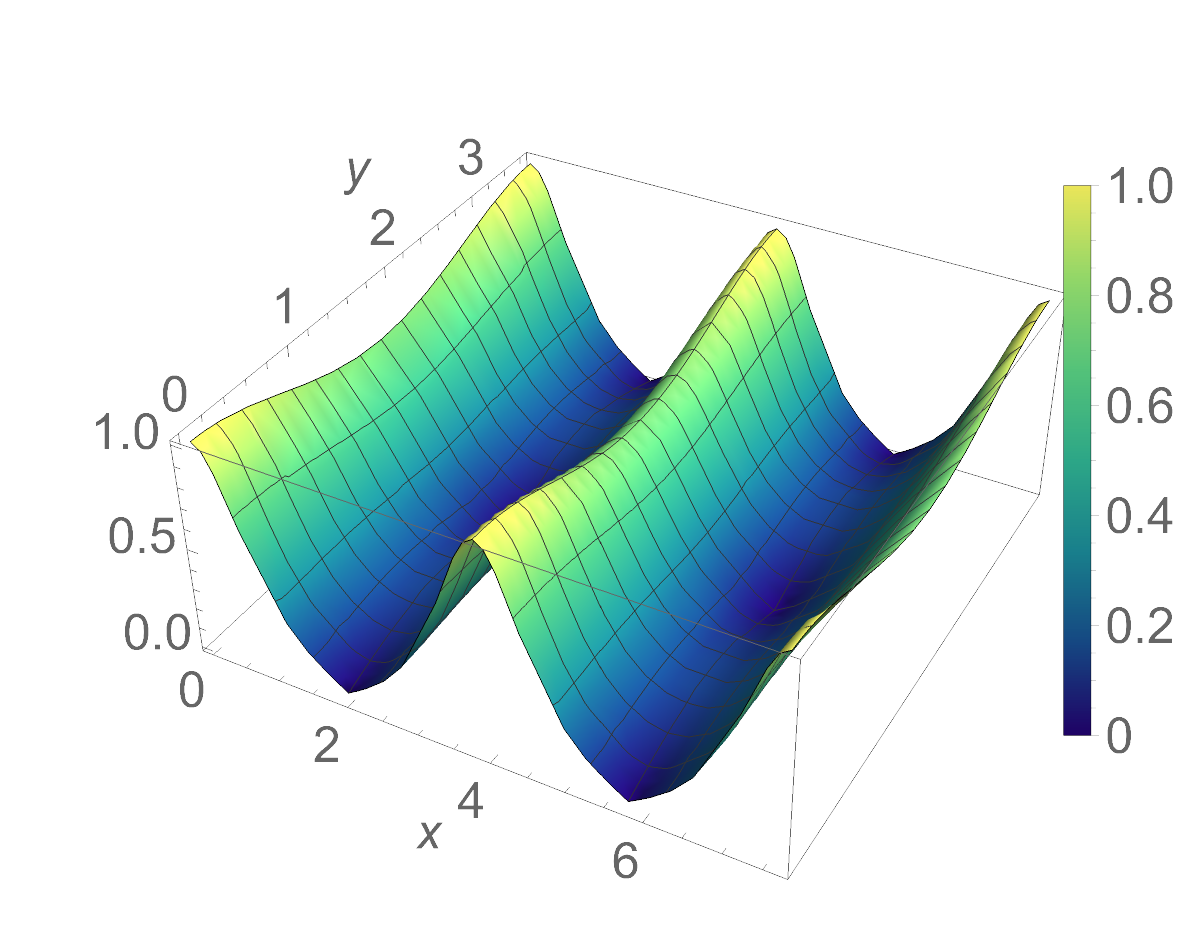}
    \includegraphics[width=0.3\textwidth]{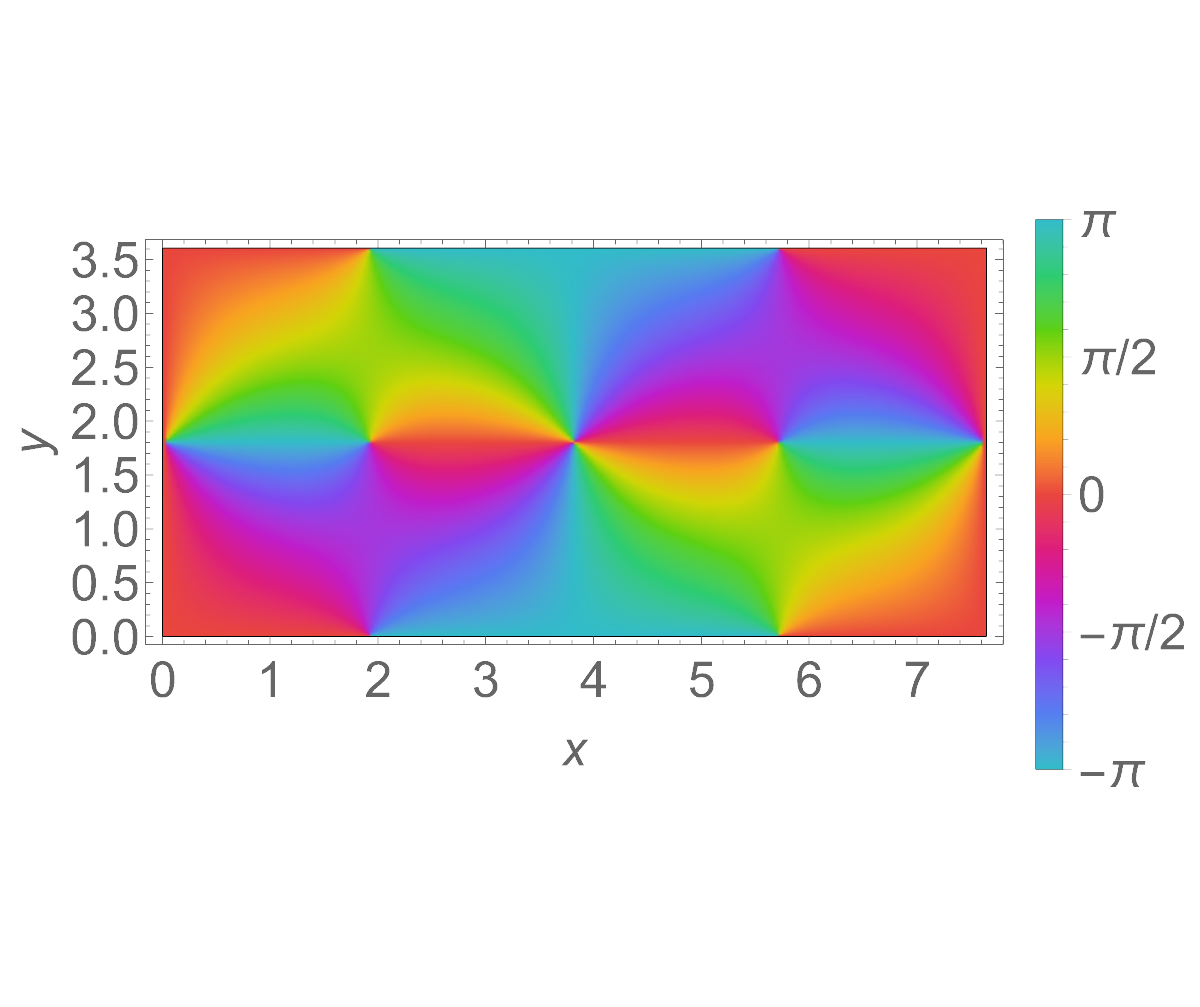}
    \caption{Profiles of $\abs{\phi_{\sn}}$ (left column) \revise{and its phase angle (right column)} with modulus $k=1/4$ (top), $k=1/2$ (middle) and $k=3/4$ (bottom).
    The domain of these plots is the fundamental lattice spanned by $4K$ and $2iK'$.}
    \label{fig:phi-sn-3D}
\end{figure}

The profiles of $\phi_\sn$ with various values of $k$ are shown in Figure \ref{fig:phi-sn-3D}.
The four simple zeroes appear, for which the phase angle rotates once around them as expected from \eqref{eq:multiple zero}.
On the other hand, the phase angle rotates twice around the two saddle points, which is consistent with \eqref{eq:Higgs zero from poles} with $n=1$, namely, the behaviour is
\begin{align}
    \phi_\sn(\mbox{saddle pts.})\sim -\frac{1}{\overline{c}}e^{-2\theta i}+O(\epsilon).
\end{align}
The vortex number of these solutions is\revise{, therefore,} four because all four zeroes are simple, which \revise{also}agrees with the numerical integration.

We note that if the phase factor of the Higgs field is ignored the fundamental domain of the solution would be halved in the real axis \revise{as  indicated by the absolute value plots of the Higgs field in Figure \ref{fig:phi-sn-3D}.}
This can be understood by the periodicity of the Jacobi elliptic functions $\sn(z+2K)=-\sn(z),\, \cn(z+2K)=-\cn(z)$, and $\dn(z+2K)=\dn(z)$ from \eqref{eq:higgs from sn}.
However, the continuity of the Higgs field as a function of $z$ and $\overline{z}$ is lost on this half-domain, on which the phase angle is not periodic at the boundaries so that the formula for the vortex number \eqref{eq:integration-F-on-Ttilde} will be invalid.
Therefore, we require the strict doubly periodicity of the Higgs field itself in our analysis for the vortices. 
We will comment on this issue again in the final example.

\paragraph{Example 3: Powers of Jacobi sn function}
For a vortex solution with multiple zeroes, we consider a solution constructed from the multiple powers of the Jacobi sn function.
\revise{As an illustration, we choose $f(z)=\sn^3(z;k)$. Then the Higgs field takes the form}
\begin{equation}\label{eq:Higgs from sn^3}
    \phi_{\sn^3}(z,\bar{z}) = \frac{3~\cn(z;k)\dn(z;k)\sn^2(z;k)}{1+\abs{\sn^3(z;k)}^2},
\end{equation}
whose fundamental lattice is the same as in \revise{the example of $\sn(z;k)$.}
\revise{This solution has three types of zeroes: 
the simple zeroes arising from $\cn(z)$ and $\dn(z)$ located at $K,3K$ and $K+iK', 3K+iK'$ respectively, the double zeroes arising from $\sn^2(z)$ located at $0$ and $2K$, and the double zeroes arising from the triple poles of $\sn^3(z)$ in the denominator located at $iK'$ and $2K+iK'$.}
We \revise{thus} observe that the vortex number of this solution is \revise{$4\cdot 1+2\cdot 2+2\cdot 2=12$.}
This is consistent with the result of numerical integration.

\begin{figure}[htbp]
    \centering
     \includegraphics[width=0.5\textwidth]{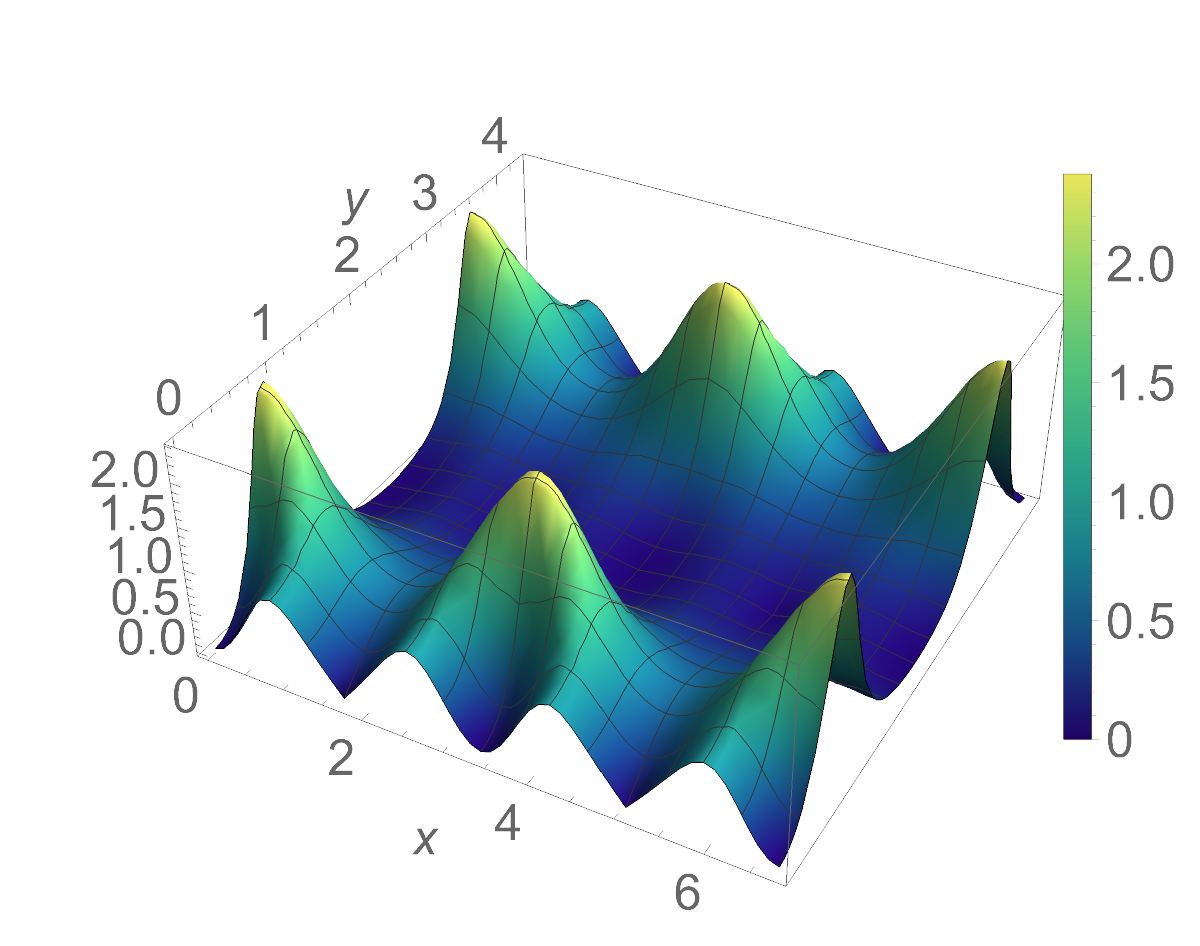}
      \includegraphics[width=0.4\textwidth]{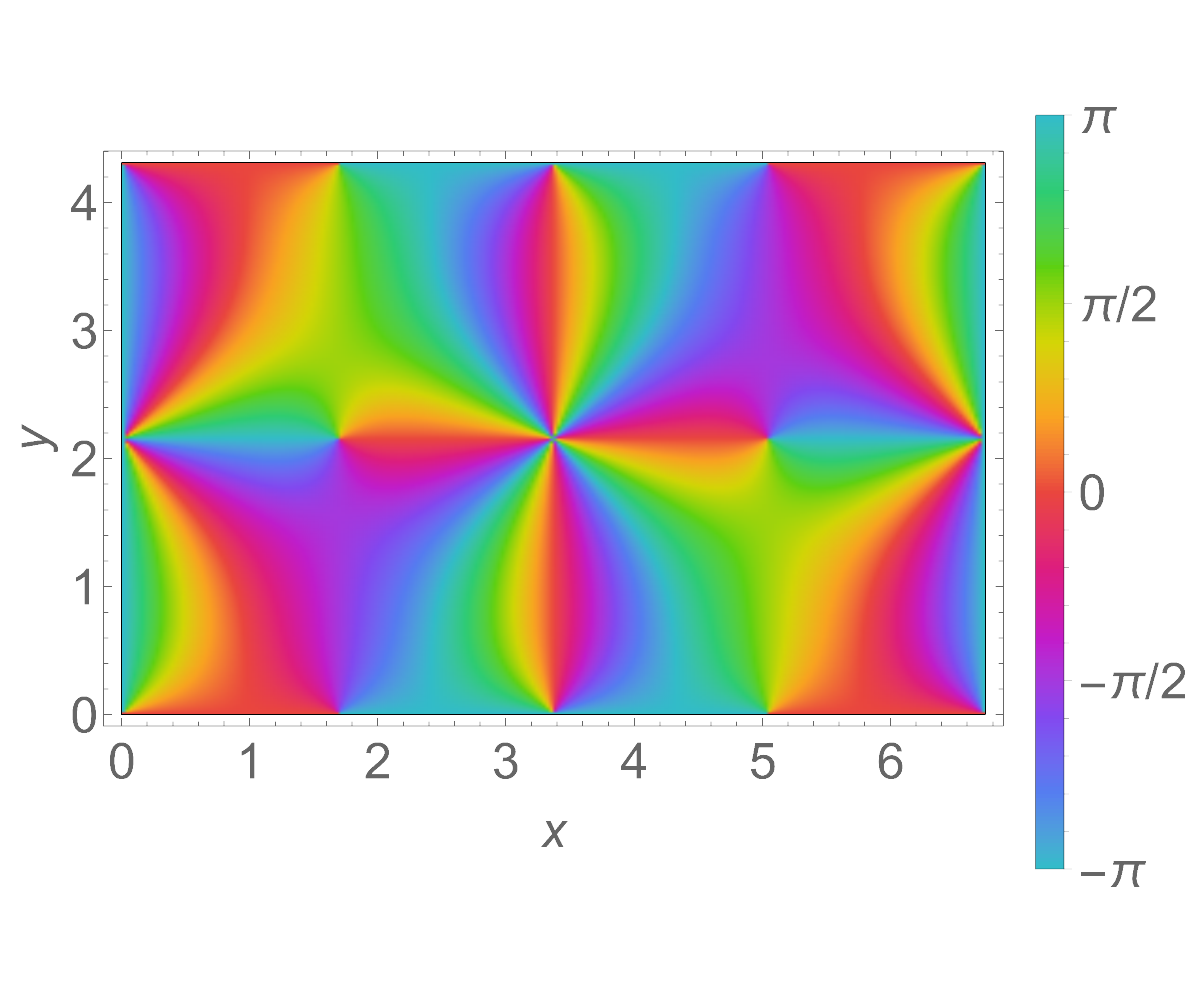}
    \caption{Profile of $\abs{\phi_{\sn^3}}$ (left) \revise{and its phase angle (right). the modulus $k=1/2$, and the fundamental region is spanned by $4K$ and $2iK'$.}}
    \label{fig:phi-sn-3powers-3D}
\end{figure}

Figure \ref{fig:phi-sn-3powers-3D} shows the profile of this solution \revise{and its phase.}
The simple zeroes are located at $K,3K,K+iK'$ and $3K+iK'$ with phase angle rotation $2\pi$.
The double zeroes from the numerator at $0$ and $2K$ show the phase angle rotation twice, i.e., $4\pi$ as expected from \eqref{eq:multiple zero} with $n=2$, whereas those from the poles of the denominator at $iK'$ and $2K+iK'$ demonstrate the rotation angle $8\pi$ as expected from \eqref{eq:Higgs zero from poles} with $n=3$.

\paragraph{Example 4: Concerning Olesen's solution}
\revise{In the early 1990s, Olesen constructed a Jackiw-Pi vortex on a torus with unit vortex number \cite{Olesen:1991dg} in the context of the Chern-Simons-Higgs theory.
Let us glance at discussions about this solution.
Because of the Riemann-Hurwitz formula, the vortex number of the Jackiw-Pi vortices on a torus is an even integer \cite{Manton:2016waw}.
However, the vortex number of Olesen's solution equals $1$.
This contradiction is resolved in \cite{Akerblom:2009ev} by introducing the ``$\Omega$-quasi elliptic functions", whose flux density $\rho=|\phi|^2$ is doubly periodic in the quarter cell of the fundamental lattice while the Higgs field itself is not so.
This oddness is also discussed by using a generalized Riemann-Hurwitz formula \cite{Walton:2021urz}.
Nevertheless, we consider in this paper the Jackiw-Pi vortices on $T^2$ constructed from the strict doubly periodic Higgs field $\phi(z,\bar{z})$ itself.
The reason is that the continuity of the Higgs field as a complex function is necessary for the analytic calculation of the vortex number.}

\begin{figure}[htbp]
    \centering
     \includegraphics[width=0.5\textwidth]{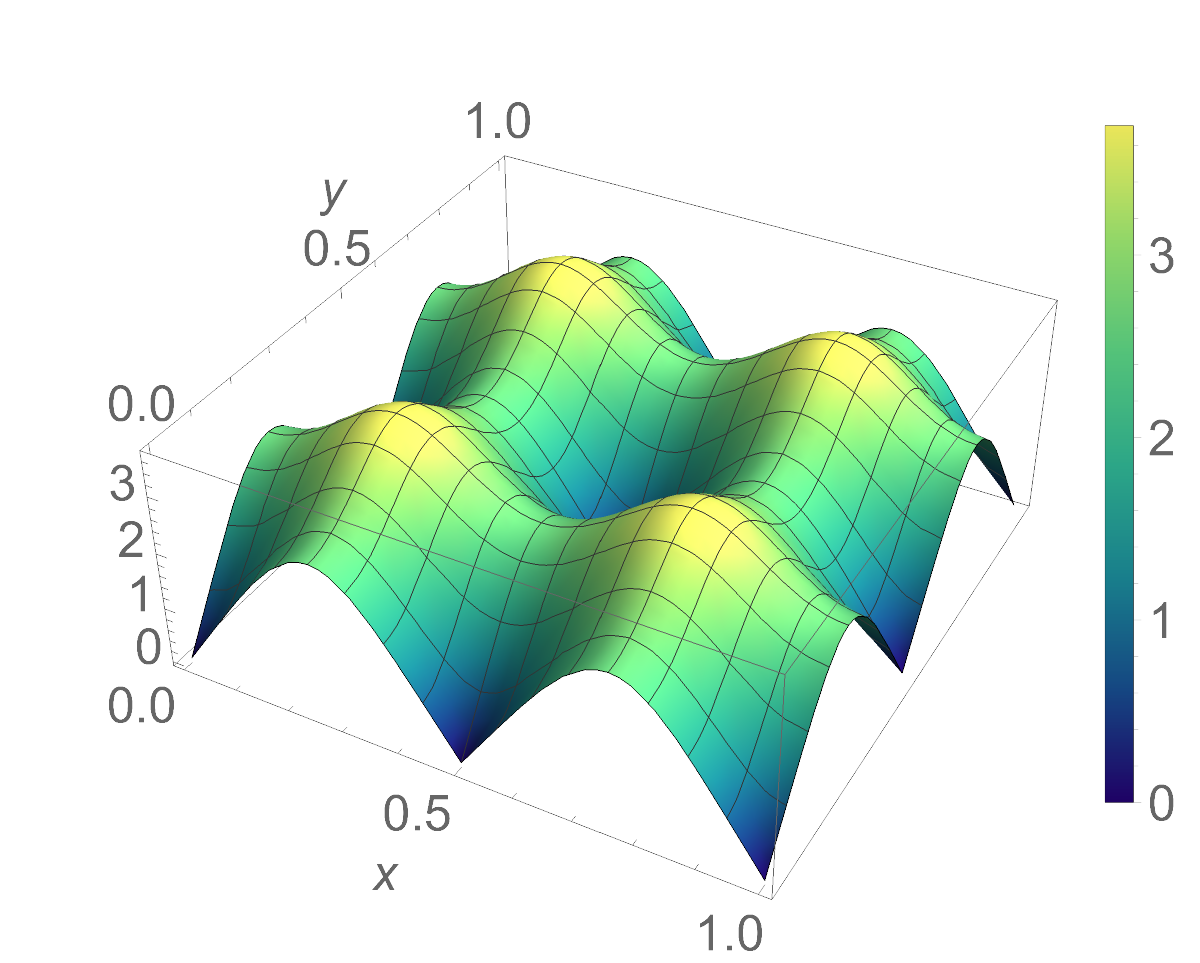}
     \includegraphics[width=0.4\textwidth]{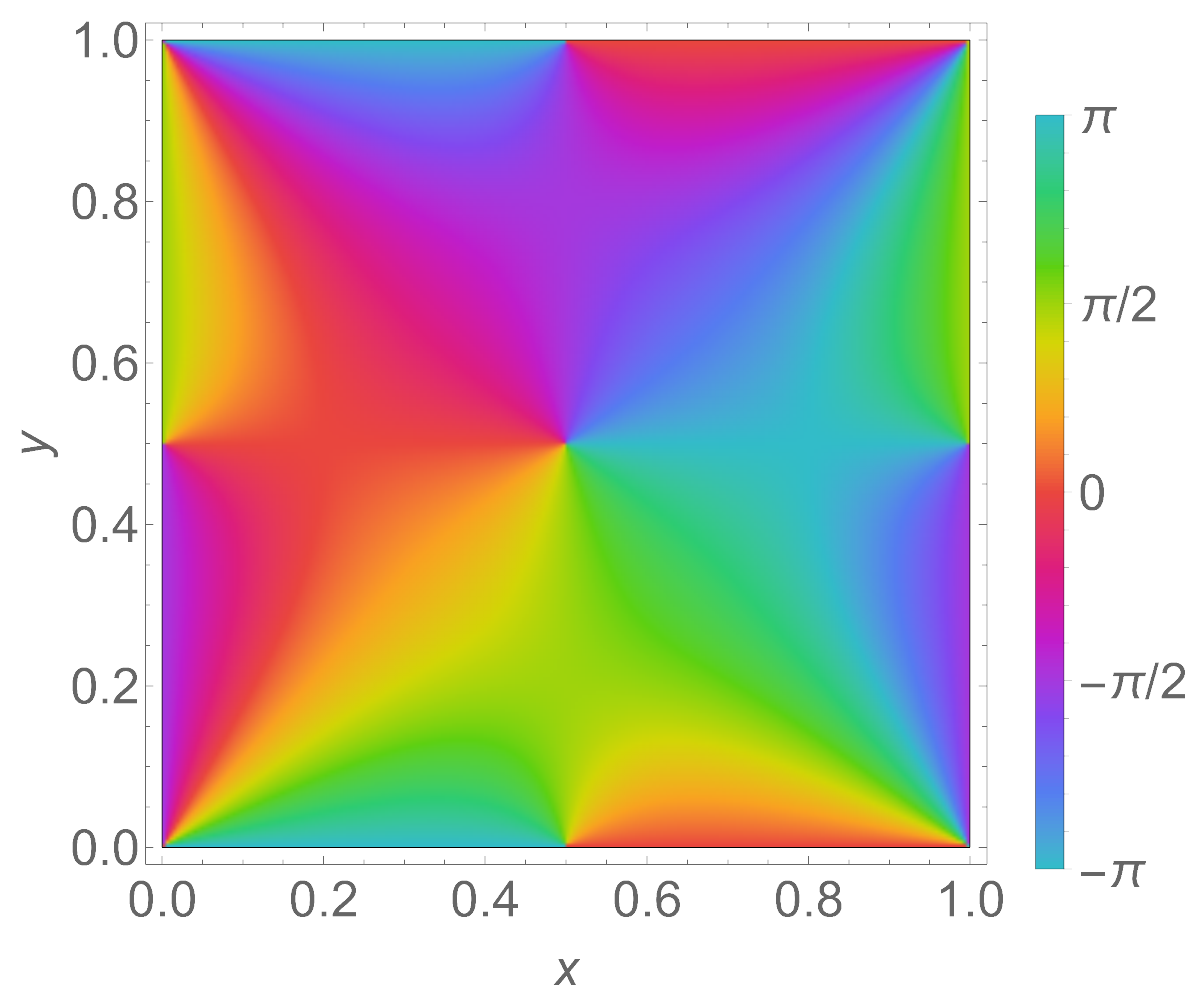}
    \caption{Profile of Olesen's solution of the square fundamental region.
    The left is the profile of $|\phi(z,\overline{z})|$ and the right is \revise{its phase angle.}
    The colour scale legend is the same as the previous Figures.}
    \label{fig:Olesen}
\end{figure}

\revise{The meromorphic function defining the solution is 
\begin{align}\label{eq:Olesen's f}
    f(z)=\frac{\wp(z)-e_3}{\sqrt{(e_3-e_1)(e_2-e_3)}}.
\end{align}
Although the solution is similar to the first example \eqref{eq:Higgs from pe}, a constant shift and an overall scaling are assembled for $\wp(z)$.
This adjustment makes the period of the flux density $\rho$ into a quarter of the fundamental region of $f(z)$, meaning that $|\phi(z+\omega_j,\overline{z+\omega_j})|=|\phi(z,\overline{z})|,\ (j=1,2,3)$.
For instance, we show a profile of Olesen's solution for the half periods $\omega_1=1/2,\,\omega_2=i/2$, for which $e_1=-e_2\simeq 6.87519$ and $e_3=0$ in Figure \ref{fig:Olesen}.}

\revise{
The flux density of Olesen's solution is doubly periodic about the quarter cell, and numerical calculation shows the vortex number in the cell is one. 
However, from the right of Figure \ref{fig:Olesen}, we observe obviously that the phase angle of the Higgs field of the solution is not periodic in the quarter cell.
Thus our analytic calculation is inapplicable for the solution in such a smaller domain. 
We, hence, should apply the whole fundamental domain for $f(z)$ as the fundamental lattice of the solution,
and the vortex number of this solution is $4$.
This understanding will be acceptable from the point of view of the Aharonov-Bohm-like effects, in which the phase angle of fields plays a critical role.}

\section{Large period limits of vortices}

If we take a fundamental period of the elliptic functions to be infinite, the fundamental lattice expands in that direction and the torus turns out to be a cylinder.
Furthermore, the fundamental lattice becomes planar if both of the periods are set to be infinite.
In this section, we consider the vortex solutions defined on such large period limits of the fundamental lattice.
We will find characteristic ``flux loss" phenomena of vortices in those limiting cases.

\subsection{Cylinder limits of the vortex from Jacobi elliptic function}
To consider the cylinder limit, the vortex made from the Jacobi elliptic functions is favourable because they become elementary functions in the limits.

\begin{figure}[h]
    \centering
    \includegraphics[width=0.19\textwidth]{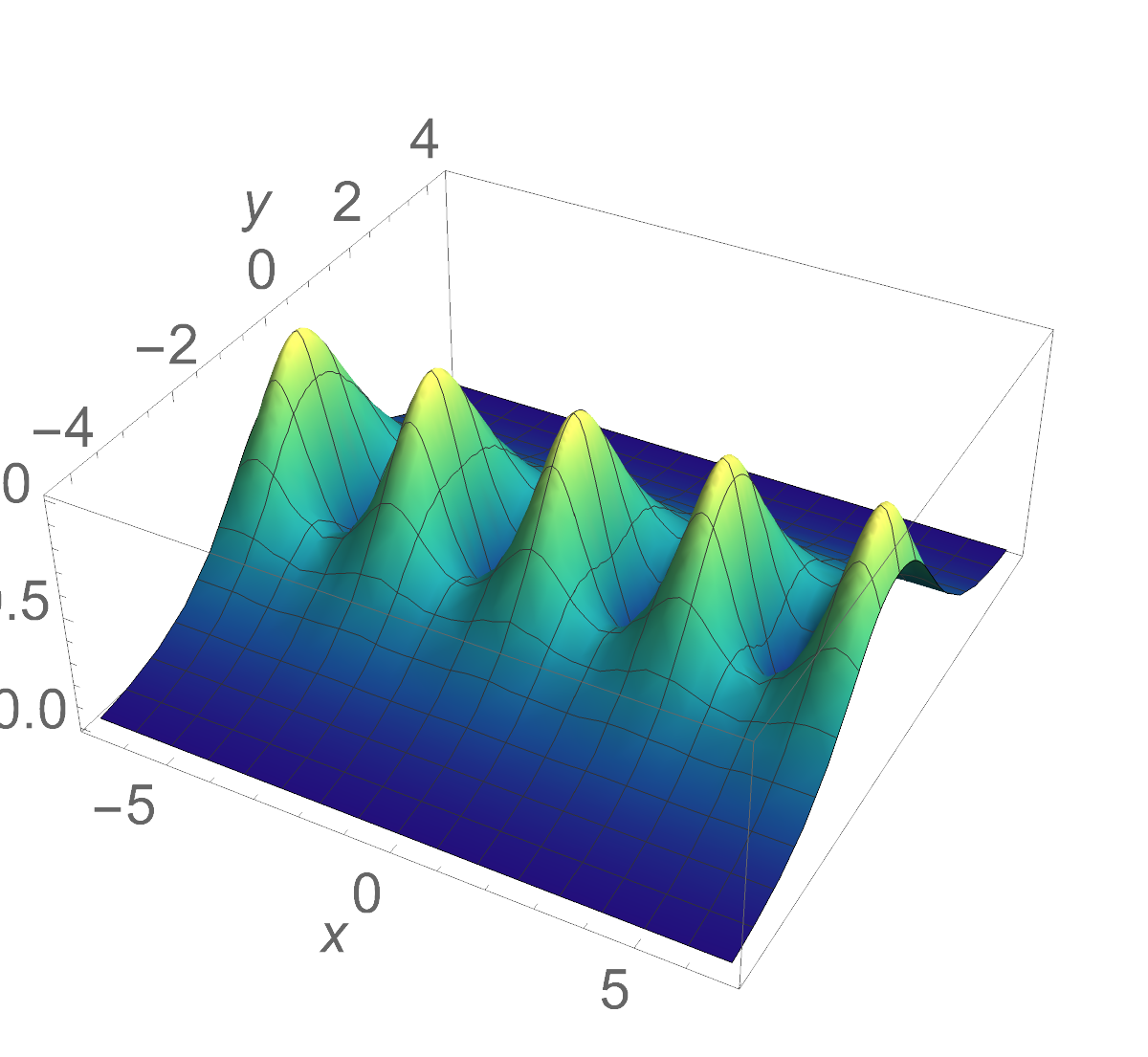}
    \includegraphics[width=0.19\textwidth]{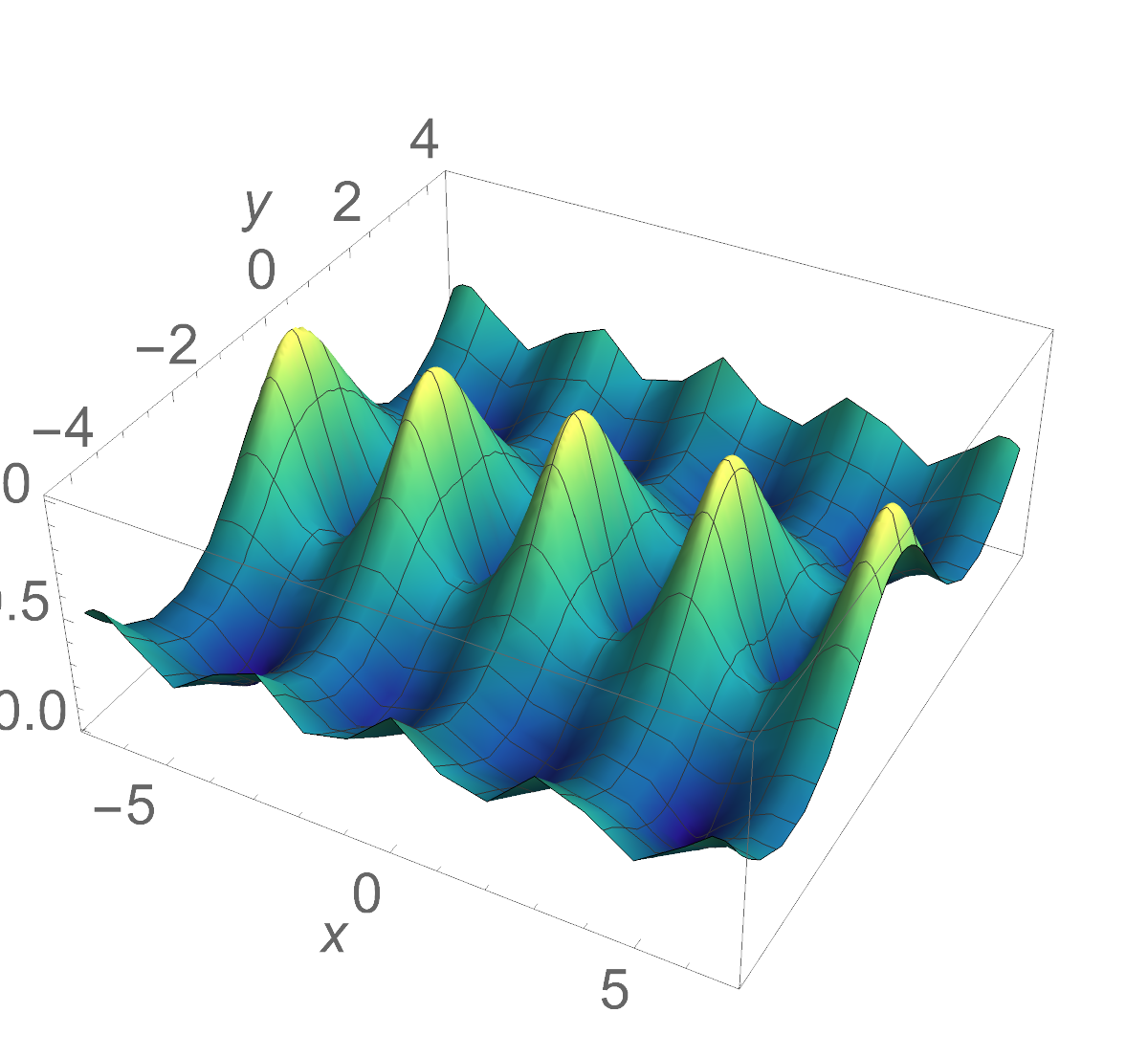}
    \includegraphics[width=0.19\textwidth]{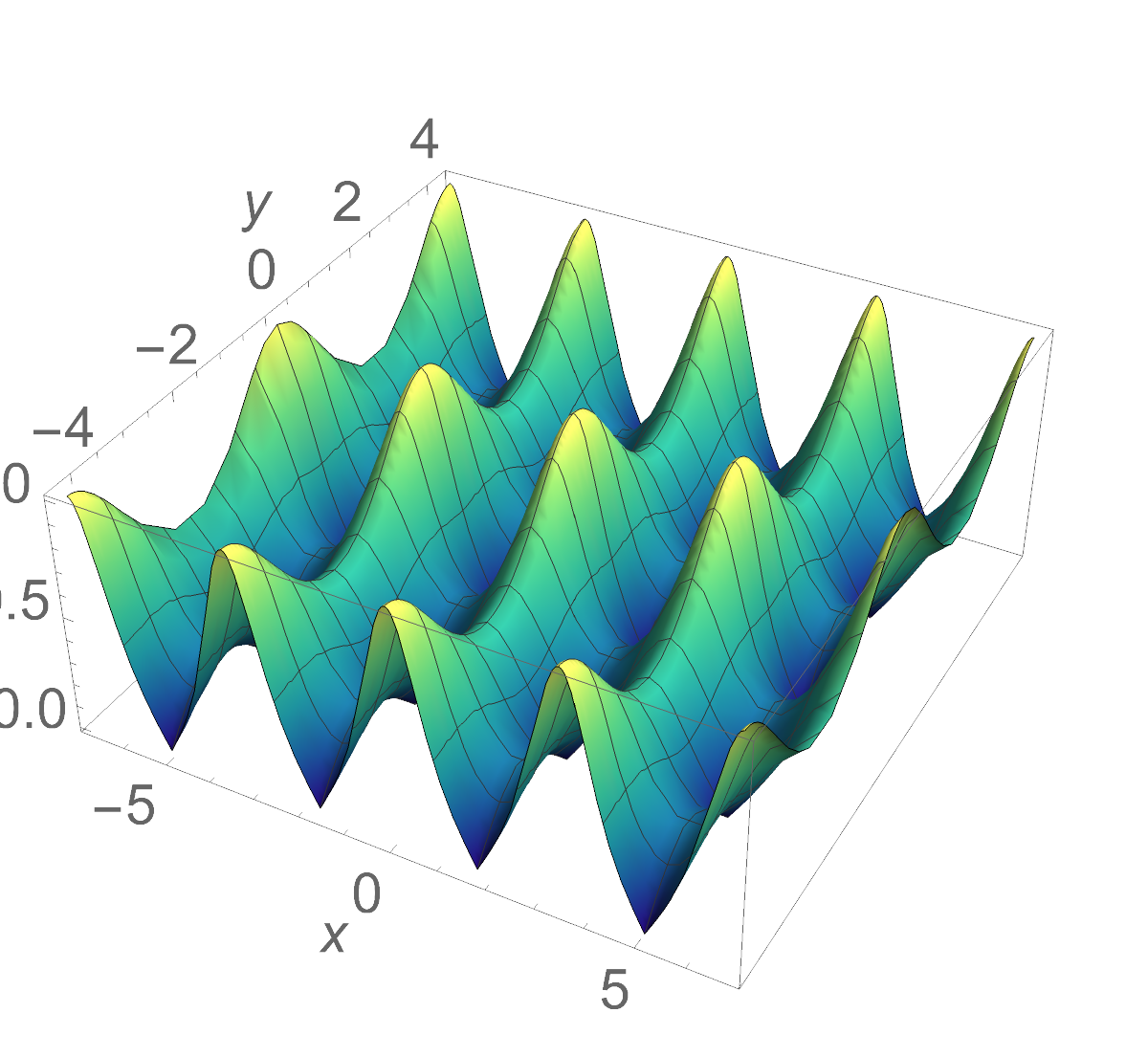}
    \includegraphics[width=0.19\textwidth]{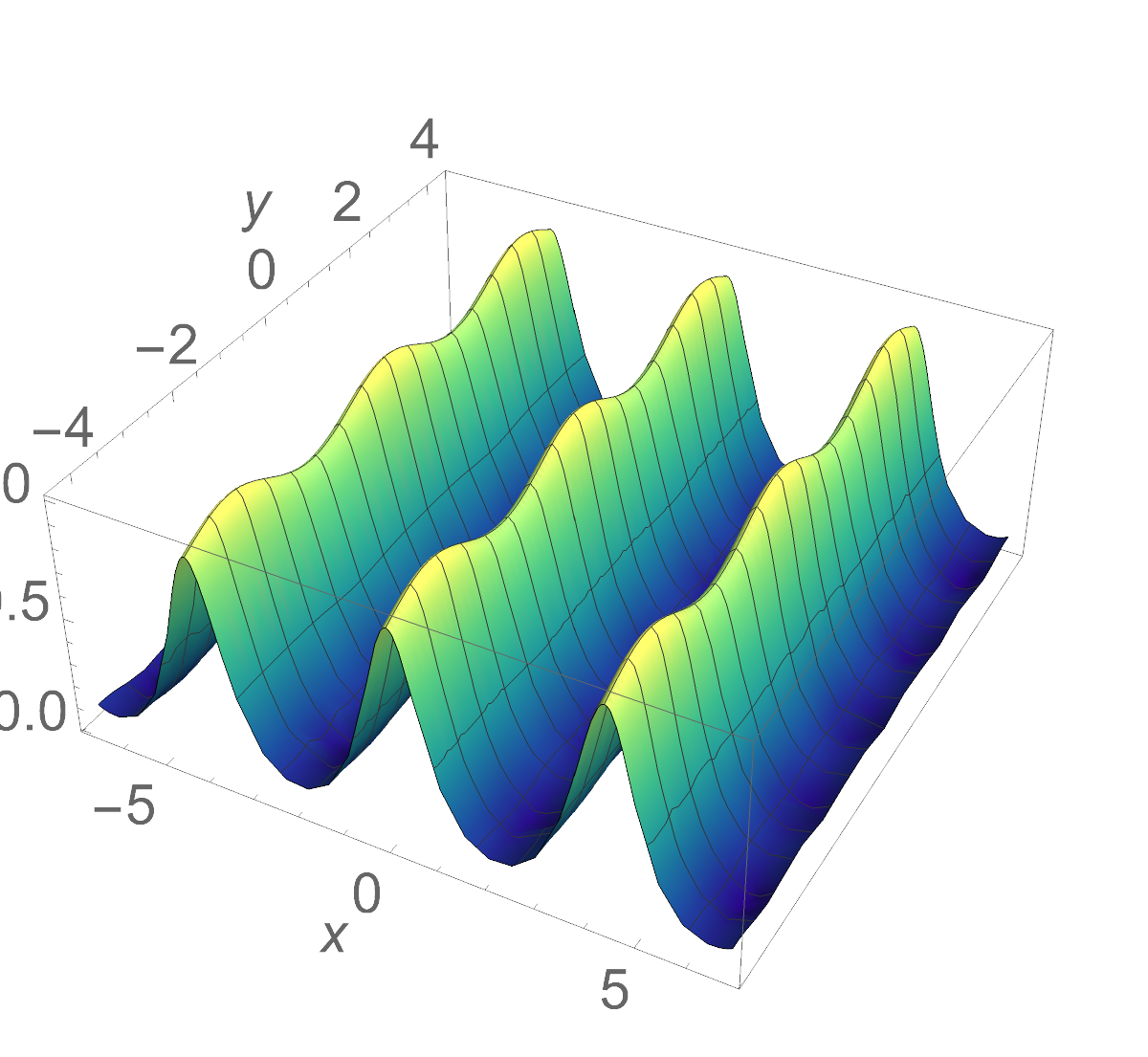}
    \includegraphics[width=0.19\textwidth]{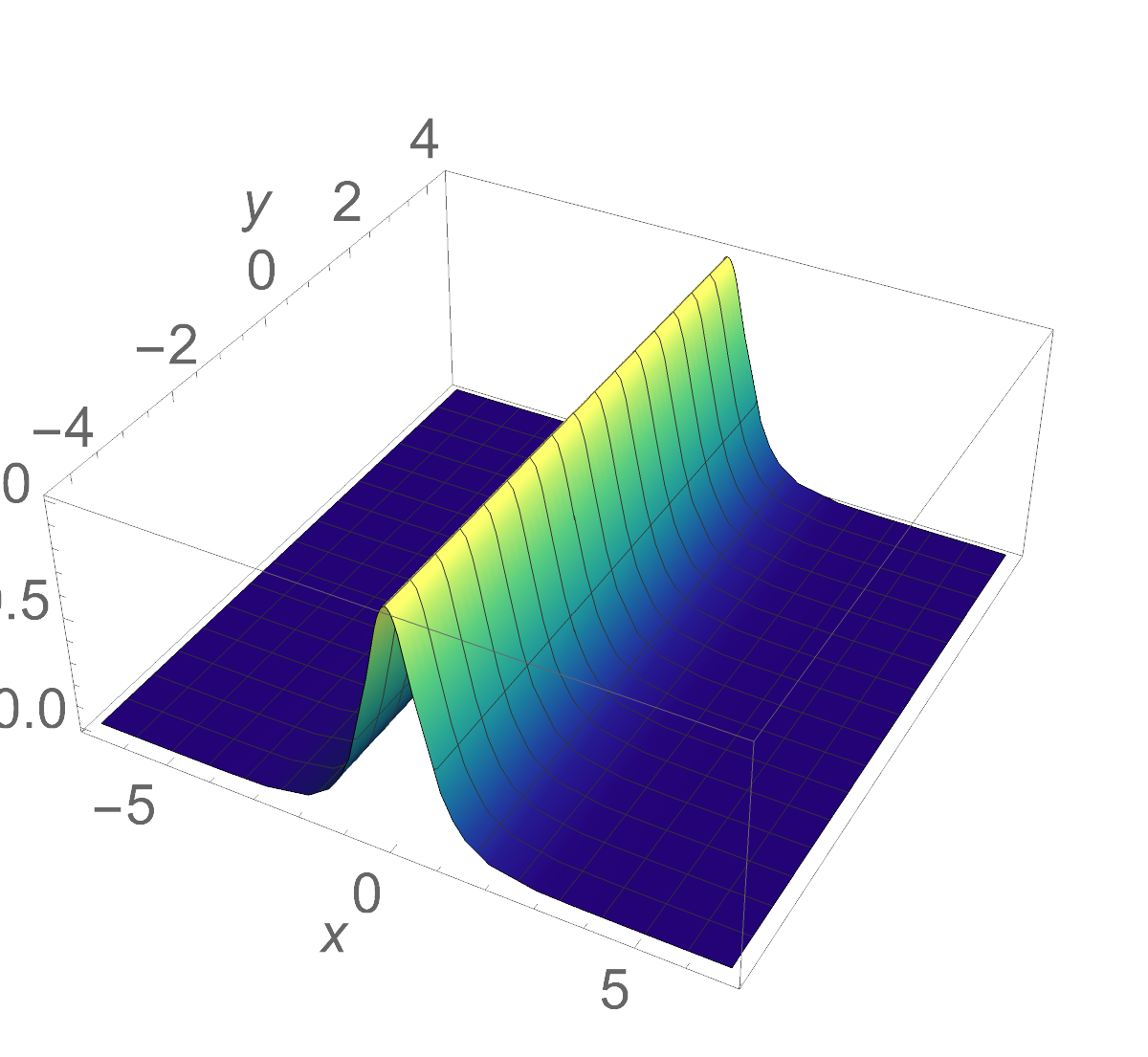}
    \caption{\revise{Changes in the profile of $\abs{\phi_{\sn}}$ as the modulus $k$ varies. The values of $k$ are $0, 0.25,0.5,0.85$, and $1$, from left to right, respectively. Each profile is plotted within the same rectangular region $-4K(k)\leq \Re(z) \leq 4K(k),\ -4K'(k) \leq \Im(z) \leq 4K'(k)$ with $k=0.5$.}}
    \label{fig:sn-cylinder-series}
\end{figure}

We reconsider the second example in the last section, i.e., the case of $f(z)=\sn(z)$ with a modulus $k$.
\revise{The behaviour of the Higgs field as $k$ varies is shown in Figure \ref{fig:sn-cylinder-series}.}
The cylinder limits are obtained by taking the limit $k\to0$ or $1$, for which $\sn(z)\to\sin(z)$ or $\tanh(z)$, respectively.
Here we consider the profiles of the Higgs field in these limits and calculate the vortex number of such cases.
Then we confirm that they are consistent with numerical integration.

\begin{figure}[h]
    \centering
    \includegraphics[width=0.5\textwidth]{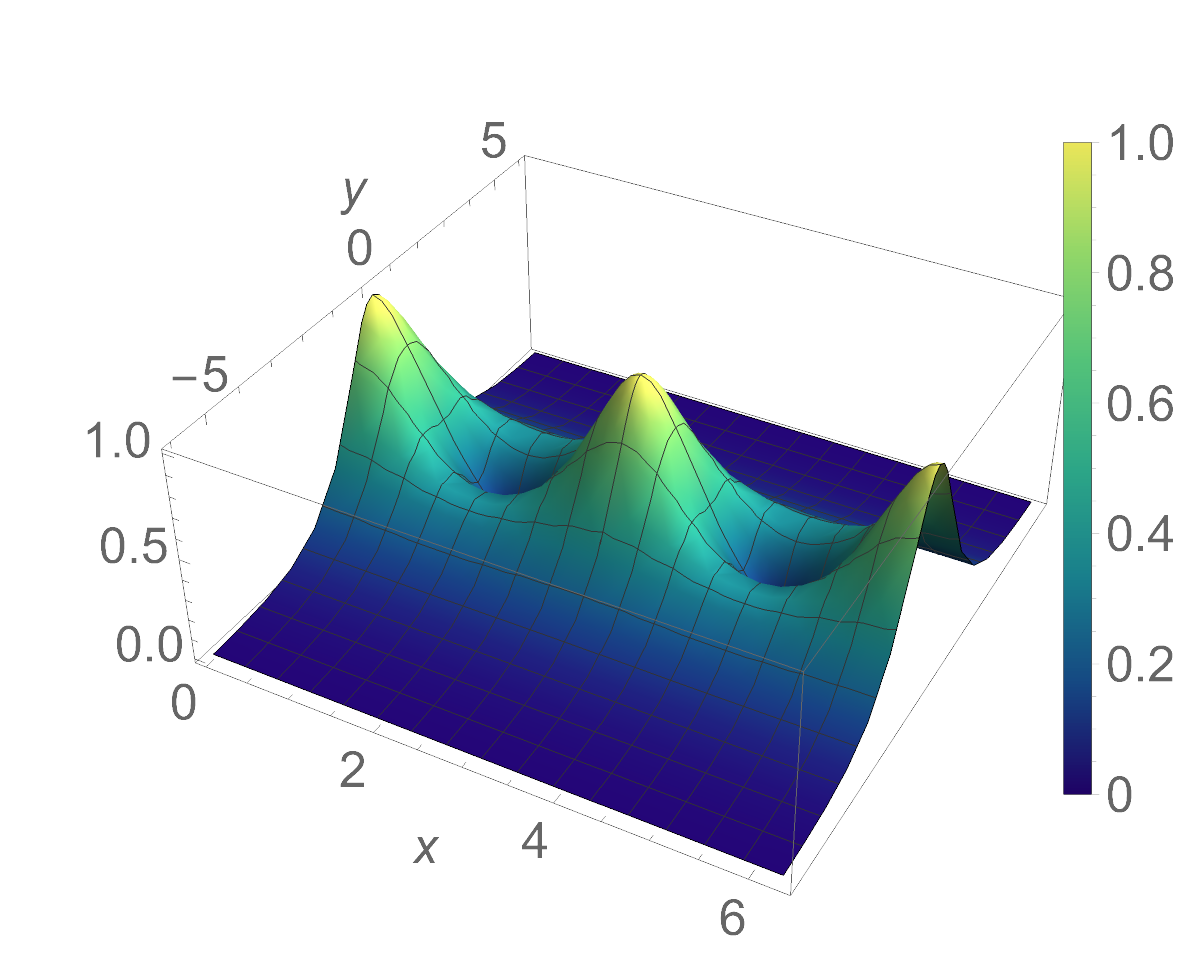}
    \includegraphics[width=0.4\textwidth]{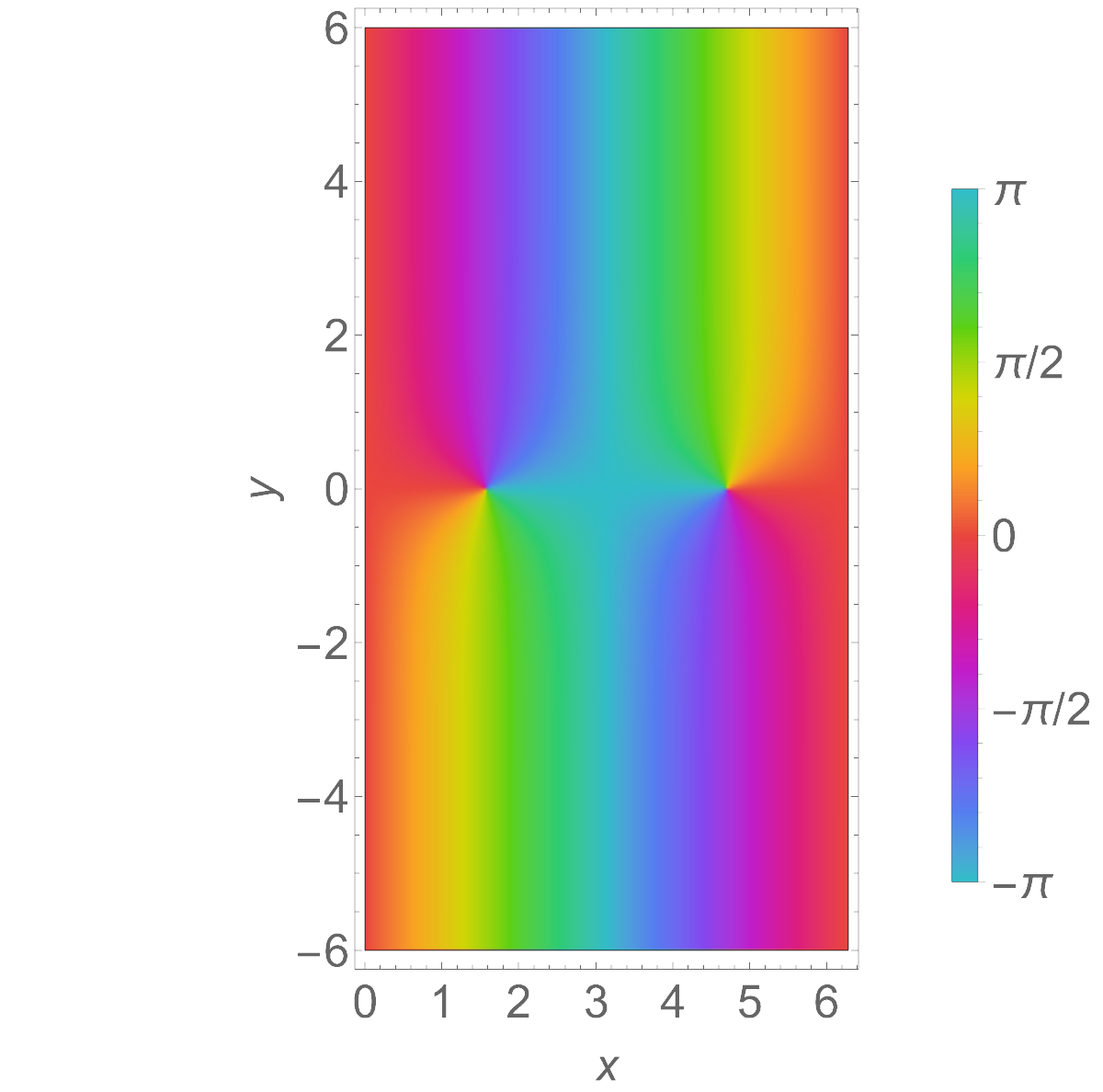}
    \caption{Profile of $\abs{\phi_{\sin}}$ (left) \revise{and its phase angle (right).}
    The periodicity persists along the real axis, \revise{but} vanishes along the imaginary axis. 
    Hence the fundamental region is an infinitely long strip that contains two zeroes.}
    \label{fig:sn-cylinder-m0}
\end{figure}

Firstly, we consider the trigonometric function limit $k\to0$.
The Higgs field $\phi_{\sn}$ then becomes
\begin{align}
    \phi_{\sn}(z,\overline{z}) \xrightarrow[k\to0]{} \phi_{\sin}(z,\overline{z}) = \frac{\cos z}{1+\abs{\sin z}^2},
\end{align}
which has simple zeroes at $z=(2m+1)\pi/2$ with $m\in\mathbb{Z}$.
The fundamental region of $\sin z$ has an infinite period in the imaginary direction and a period $2\pi$ in the real direction, as shown in Figure \ref{fig:sn-cylinder-m0}.
As a result, the point at infinity is excluded from the fundamental region, in other words, the region becomes open and can be thought of as a cylinder.
We parametrise the fundamental region of $\sin z$ with $z=x+iy$ as $0\leq x\leq 2\pi$ and $-\infty<y<\infty$, for which the simple zeroes of $\cos z$ are located at $z=\pi/2$ and $3\pi/2$.

The numerical integration for the vortex number in this cylinder limit is $4$ as in the case of a general value of \revise{$0<k<1$.}
We now derive this vortex number analytically through \eqref{eq:integration-F-on-Ttilde}, in that the integration along boundaries at infinity contributes to the vortex number.
We evaluate the contour integral
\begin{align}\label{eq:vortexnumber-integral-kto0}
    N_{k\to0} = \frac{1}{2\pi}\int_{\widetilde{\mathrm{Cyl}}} F_{z\bar{z}} dz\wedge d\bar{z} = \frac{1}{4\pi i}\int_{\pd\;\widetilde{\mathrm{Cyl}}} \pd\log\abs{\phi_{\sin}}^2 dz - \overline{\pd}\log\abs{\phi_{\sin}}^2 d\bar{z},
\end{align}
where $\widetilde{\mathrm{Cyl}}$ is the cylinder with the Higgs zeroes removed as in the case of the torus.
The difference from the torus cases is that the boundary $\pd\widetilde{\mathrm{Cyl}}$ includes the edges at the infinities in \revise{the} imaginary coordinate.
Thus, the contour integral in the right-hand-side of \eqref{eq:vortexnumber-integral-kto0} is composed of the small circles around the two Higgs zeroes and the two edges at infinities.
Since the two zeroes at $z=\pi/2$ and $3\pi/2$ are simple, the integration around these zeroes contributes to 2 from \eqref{eq:multiple zero}, so that it can be expected that the rest comes from the integration along the edges $E_{y\to\pm\infty}:=\{z=x+iy\;|\; 0\leq x\leq 2\pi,\, y\to\pm\infty\}$.
We observe the integrands of \eqref{eq:vortexnumber-integral-kto0} are
\revise{
\begin{align}
    \pd\log\abs{\phi_{\sin}}^2 = -\frac{2\cos z \sin \bar{z}}{1+\abs{\sin z}^2} - \tan z, \quad \overline{\pd}\log\abs{\phi_{\sin}}^2 = -\frac{2\cos \bar{z} \sin z}{1+\abs{\sin z}^2} - \tan \bar{z},
\end{align}}
and 
\begin{align}\label{eq:limit of sin}
    \pd\log\abs{\phi_{\sin}}^2 \xrightarrow[y\to+\pm\infty]{} \pm i.
\end{align}
The limits of $\overline{\pd}\log\abs{\phi_{k\to0}}^2$ are complex conjugates of \eqref{eq:limit of sin}.
Hence, $N_{k\to0}$ can be calculated as follows.
\begin{align}
    N_{k\to0} &= \frac{1}{4\pi i}\qty{ \int_{E_{y\to+\infty}} \left(idz - (-i)d\bar{z}\right) + \int_{E_{y\to-\infty}} \left((-i)dz - (i)d\bar{z}\right) + \text{(Contribution from zeroes)}}\nonumber\\
    &= \frac{1}{4\pi i}\qty{ \int^{2\pi}_{0} 2idx + \int^{0}_{2\pi} (-2i)dx } + 2= \frac{1}{4\pi i}\qty{ 4i\times 2\pi } + 2 = 4.
\end{align}
Here the orientation of the integral at the edges is set to ensure that the circle rotates counterclockwise around zeroes at infinities, and we should interpret that the Higgs zeroes sent away to infinity still contribute to the integration.
This is consistent with numerical integration as mentioned above.

\begin{figure}[h]
    \centering
    \includegraphics[width=0.5\textwidth]{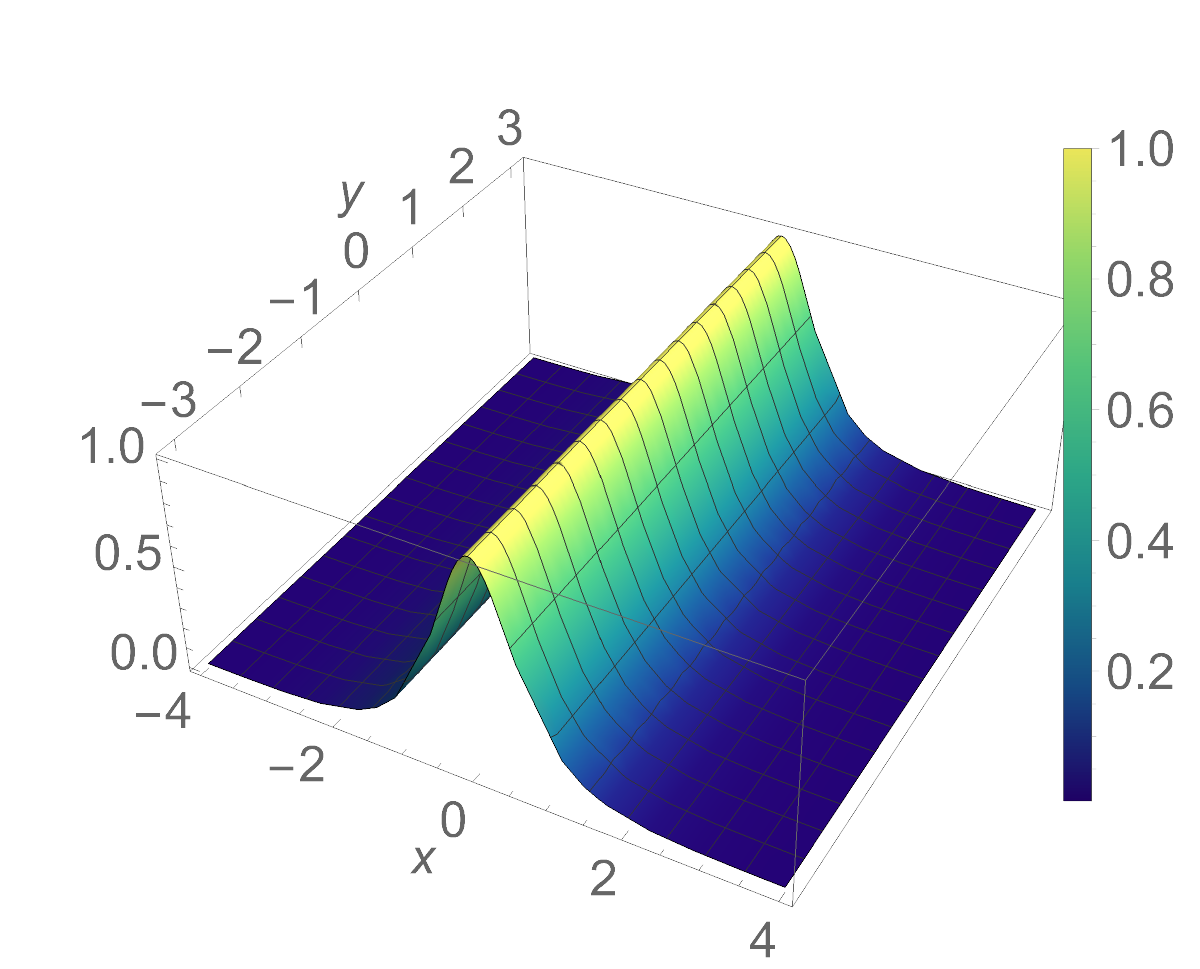}
    \includegraphics[width=0.45\textwidth]{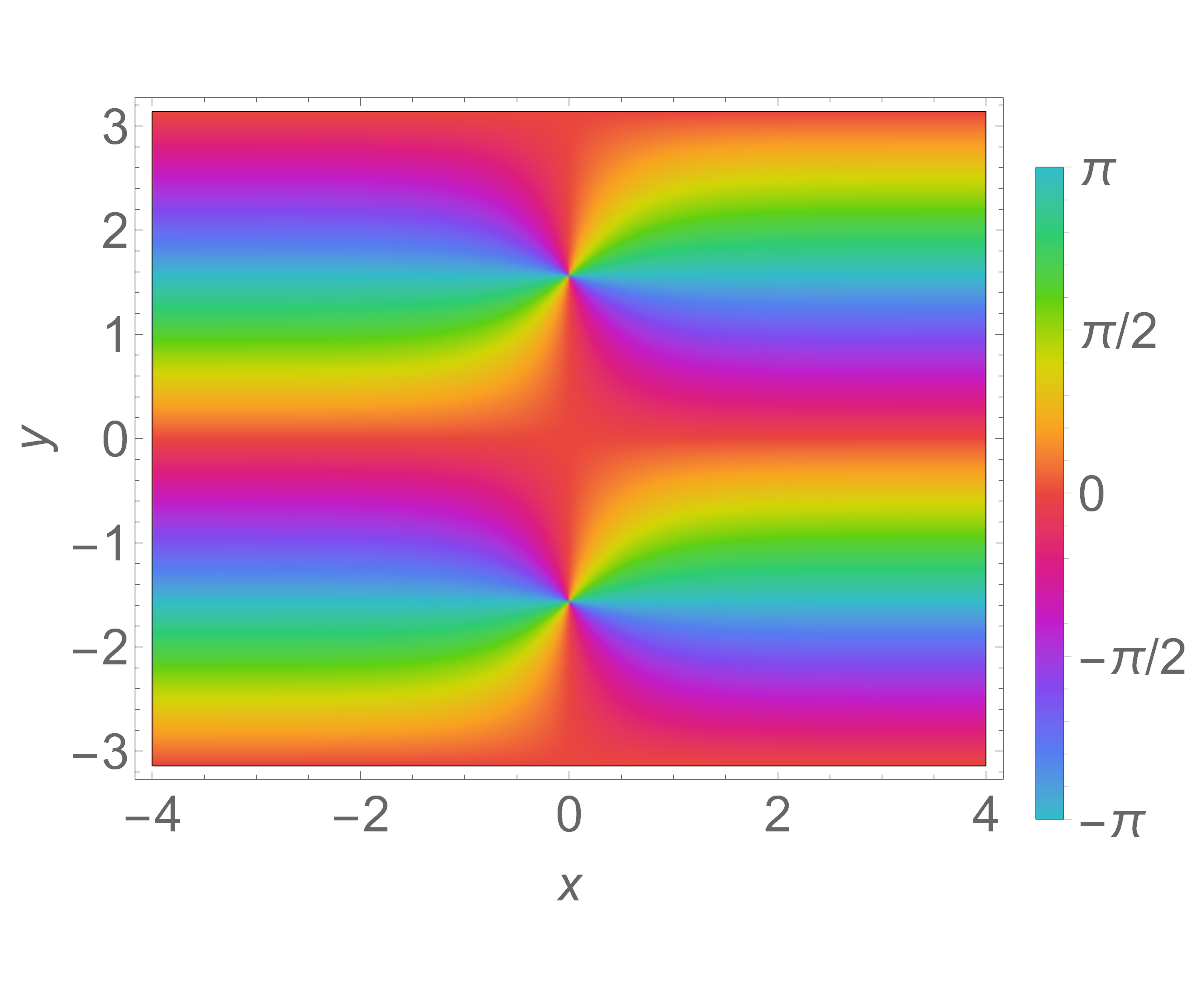}
    \caption{Profile of $\abs{\phi_{\tanh}}$ (left) \revise{and its phase angle (right).}
    The fundamental region is an infinitely long strip that contains no zeroes.
    Note that these figures are drawn for \revise{two} periods in the imaginary axis. 
    The phase angle plot shows a remnant of the periodicity along the imaginary axis.
    }
    \label{fig:sn-cylinder-m1}
\end{figure}

We consider next the hyperbolic function limit $k\to1$, for which $\sn(z;k)$ turns into $\tanh z$.
Then the Higgs field becomes
\begin{align}
    \phi_{\sn}(z,\overline{z}) \xrightarrow[k\to1]{} \phi_{\tanh}(z,\overline{z})= \frac{\sech^2(z)}{1+\abs{\tanh(z)}^2}.
\end{align}
In contrast to the former case, the fundamental region of this function has an infinite period in the real direction and a period $2\pi$ in the imaginary directions, as shown in  Figure \ref{fig:sn-cylinder-m1}.
Hence the fundamental region is also considered as a cylinder.
We take the parametrisation of the fundamental region $z=x+iy$ as $-\infty< x<\infty$ and $0\leq y\leq \pi$.

The numerical integration for the vortex number in this limit is $2$, in contrast to the trigonometric function case.
In the fundamental region, $\phi_{\tanh}$ has no zeroes, thus \revise{the contributions} to the vortex number of $\phi_{\tanh}$ will come from the integration on the edges of the region.
We now give the analytic calculation for this integration similarly to the former.

The vortex number $N_{k\to1}$ is also given by \eqref{eq:integration-F-on-Ttilde},
\begin{align}    
    N_{k\to1} = \frac{1}{4\pi i}\int_{\pd\; \widetilde{\mathrm{Cyl}}} \pd\log\abs{\phi_{\tanh}}^2 dz - \overline{\pd}\log\abs{\phi_{\tanh}}^2 d\bar{z}
\end{align}
where $\pd\; \widetilde{\mathrm{Cyl}}$ are only the edges of the cylinder $E_{x\to\pm\infty}:=\{z=x+iy\;|\; x\to\pm\infty,\;0\leq y\leq\pi$ in this case.
The integrands take the following form
\begin{align}
    \pd\log\abs{\phi_{\tanh}}^2 = \overline{\pd}\log\abs{\phi_{\tanh}}^2 = -2\tanh (z+\bar{z}),
\end{align}
thus, at the edges $x\to\pm\infty$, 
\begin{align}
    -2\tanh (z+\bar{z}) &\to \mp 2.
\end{align}
Hence, $N_{k\to1}$ can be evaluated as,
\begin{align}
    N_{k\to1} &= \frac{1}{4\pi i}\qty{ \int_{E_{x\to\infty}} (-2)dz - (-2)d\bar{z} + \int_{E_{x\to-\infty}} 2dz - 2d\bar{z}}\nonumber\\
    &= \frac{1}{4\pi i}\qty{ -4\int^{0}_{\pi} idy + 4 \int^{\pi}_{0} idy } 
    = \frac{1}{4\pi i}\qty{ 4i\times 2\pi } = 2.
\end{align}
We should interpret here that two of the four Higgs zeroes fled away from the integration in contrast to the former case.
This is also consistent with the numerical integration and illustrates the flux loss phenomena.

\subsection{Planar limit}
In the examples of the cylinder limit considered so far, the vortex number is not conserved in the limit $k\to1$, while it remains in the limit $k\to0$.
We note that a similar flux loss phenomenon has been reported in \cite{Akerblom:2009ev}, where the authors constructed the vortex on a torus from the elliptic function
\begin{align}\label{eq:Akerblom}
    f(z) = \frac{\wp'(z;t/2,it/2)}{\wp(z;t/2,it/2)}.
\end{align}
Here $\wp(z;t/2,it/2)$ is the Weierstrass $\wp$ function defined on the fundamental latice $\Lambda = \mathbb{Z}~t+\mathbb{Z}~it$.
They have taken the planar limit $t\to\infty$ of the lattice, for which the function \eqref{eq:Akerblom} tends to $-2/z$, and shown that the vortex number in this limit equals half of that of the torus.
Although the meromorphic function $f\sim 1/z$ does not give a Higgs zero from \eqref{eq:Higgs zero from poles} as in the last example, this case also has a vortex number since the integration contour can be taken around infinity in the planar limit.

Here we illustrate another simple case of a vortex in the planar limit.
Let us consider the first example of the last section $f(z)=\wp(z;\omega_1,\omega_2)$, which leads to a vortex of the vortex number $4$ on a torus.

\begin{figure}[h]
    \centering
    \includegraphics[width=0.5\textwidth]{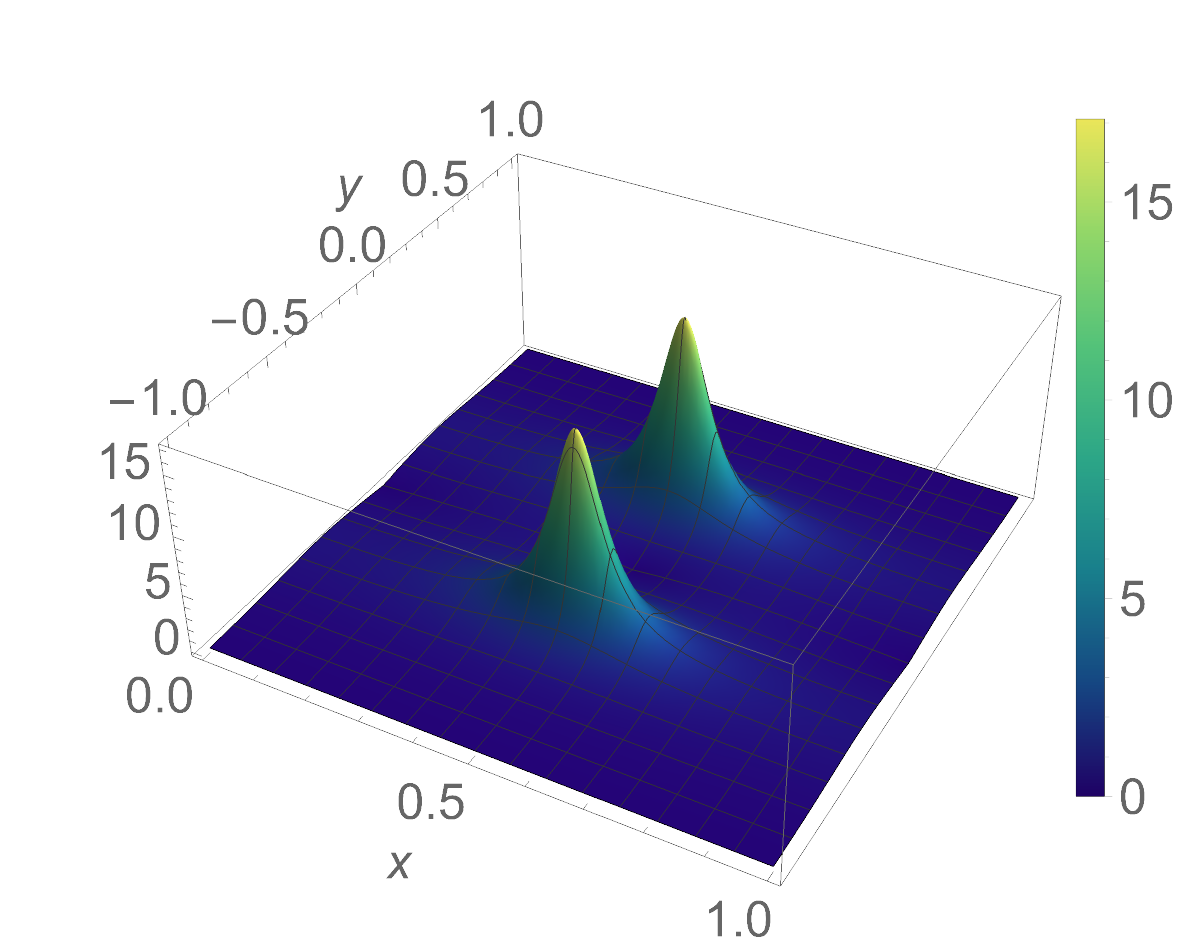}
    \includegraphics[width=0.45\textwidth]{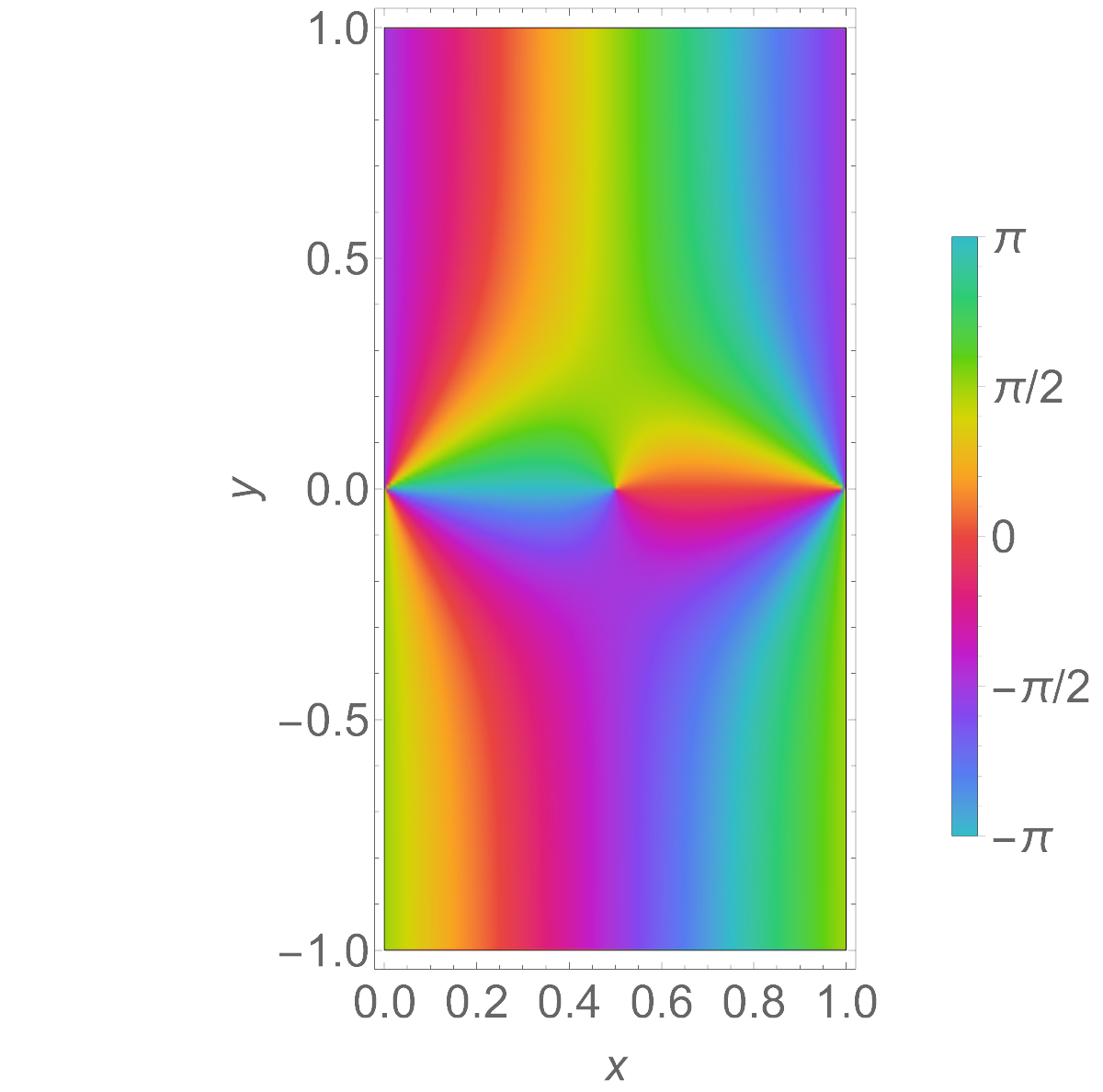}
    \includegraphics[width=0.5\textwidth]{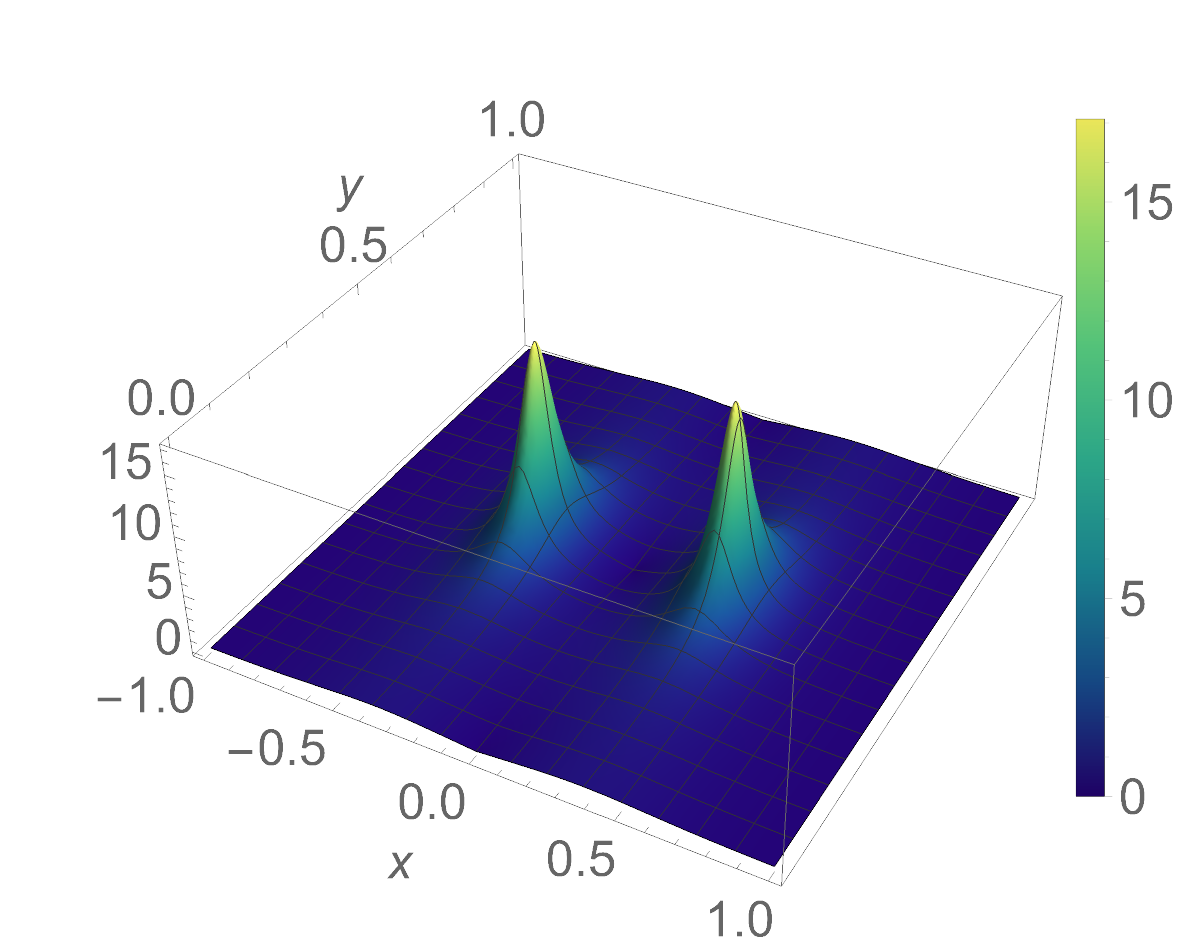}
    \includegraphics[width=0.45\textwidth]{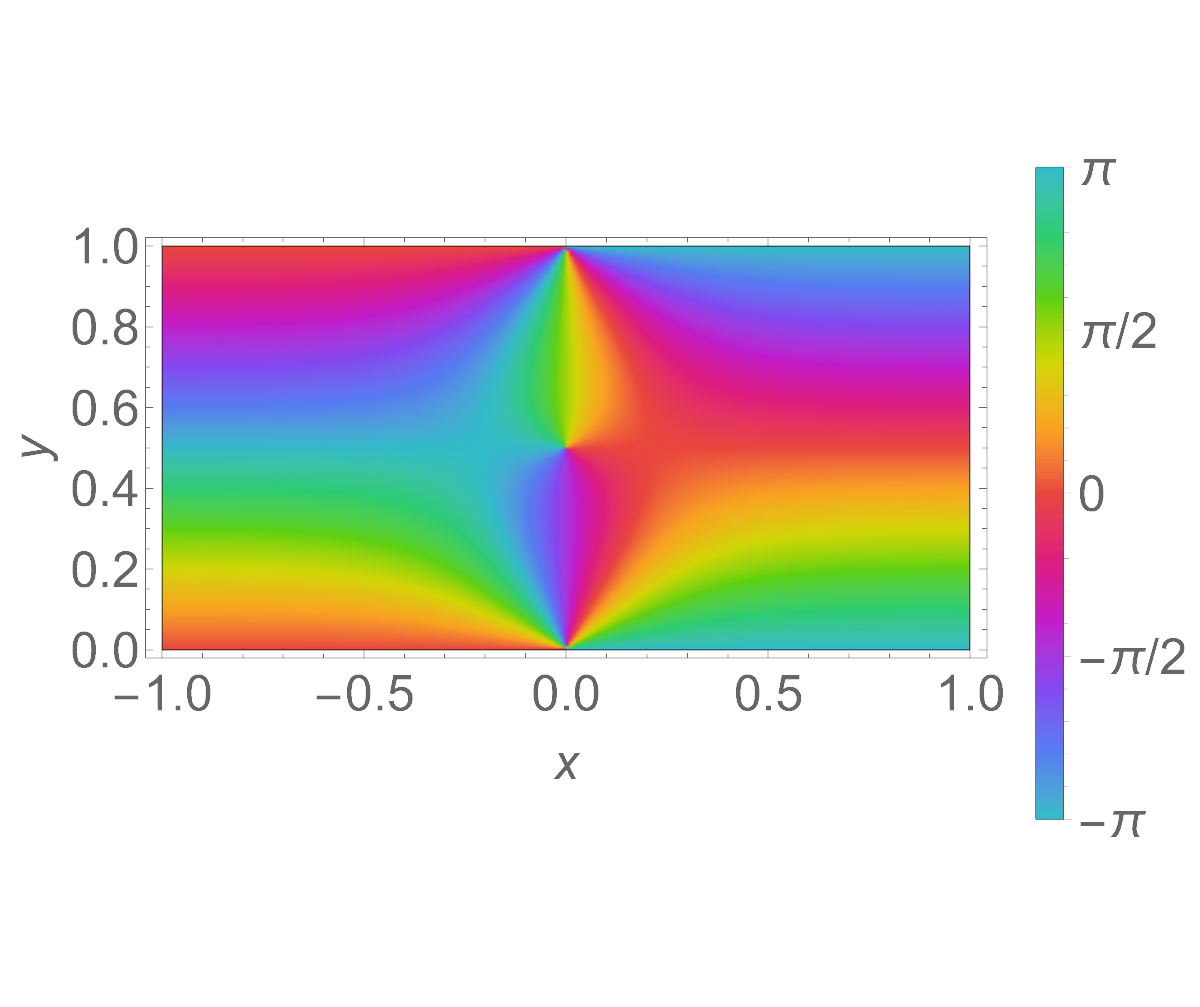}
    \caption{\revise{Profiles of the absolute value of the Higgs fields  (left column) and its phase angle (right column).
    The top row is for $\wp(z;\infty,i/2)$, in which the constants $(g_2,g_3)=(4 \pi^4 /3,-8 \pi^6 /27 )$, and the bottom row is for $\wp(z;1/2,i\infty)$, in which the constants $(g_2,g_3)=(4 \pi^4 /3, 8 \pi^6 /27 )$.}
    }
    \label{fig:sn-cylinder-Weierstrass}
\end{figure}

\revise{
Firstly, consider the cases in which one of the periods approaches infinity.
These cases are comparable to the previous cylinder limits.
Let the fundamental region be a square, then we can obtain a cylinder-like region by taking a limit $\omega_1\to \infty$.
In this limit, $f(z)=\wp(z;\omega_1,\omega_2)$ becomes
\begin{align}
    f_{\omega_1\to\infty}(z) = \wp(z;\infty,\omega_2) = \frac{1}{z^2} + \sum_{m\neq 0} \frac{1}{(z-m\omega_2)^2} - \frac{1}{(m\omega_2)^2}.
\end{align}
This function preserves the periodicity regarding the direction of $\omega_2$.
It is similar in the limit $\omega_2\to \infty$ that the periodicity regarding the direction of $\omega_1$ is preserved.
Profiles of the Higgs field in these limits are shown in Figure \ref{fig:sn-cylinder-Weierstrass}.
Numerical evaluations show the vortex number equals $4$ in both limits.
}

\revise{Secondly,} let \revise{the} both periods be infinite, then \revise{the constants $(g_2,g_3)=(0,0)$, and} the Weierstrass \revise{$\wp$} function becomes a rational function, i.e., $f(z)\to 1/z^2$ defined on a plane.
On first inspection, this vortex gives the vortex number $1$ because the meromorphic function has only a double pole at the origin from the point of view of \eqref{eq:vortex number from n-th pole}.
However, it is known that, \revise{on a flat plane,} the meromorphic functions $1/z^n$ give radially symmetric vortices of the vortex number $2n$ \cite{Horvathy:1998pe} so that the vortex number will be $4$ in this case.
We can confirm that no flux loss is observed by evaluating the flux integral as follows.
The Higgs field in the planar limit is obtained from the rational function $1/z^2$ as
\begin{align}
    \phi(z,\overline{z})=\frac{-2/z^3}{1+(1/|z|^4)}=-\frac{2\overline{z}^2/z}{1+|z|^4},
\end{align}
which is rewritten in the polar coordinates 
\begin{align} 
    \phi(z,\overline{z})=\phi(r,\theta)=-\frac{2re^{-3i\theta}}{1+r^4},
\end{align}
where $z=re^{i\theta}$ and $\overline{z}=re^{-i\theta}$.
The vortex number, or the flux integral, is easily calculated as
\begin{align}
    N_{\text{planar}}=\frac{1}{2\pi}\int_{\mathbb{R}^2}4|\phi|^2 dxdy=\frac{1}{2\pi}\int_0^{2\pi}d\theta\int_0^\infty\frac{16r^2}{(1+r^4)^2}rdr=4,
\end{align}
which indicates no flux loss in the planar limit of the vortex.
This observation shows that some relics of the vortex on a torus survive at infinity after taking the planar limit from the perspective of the plane as a decompactified torus.
The scenario would be similar to the cylinder limit of a torus considered in the last subsection.

As we have seen in this section, the vortices on a torus occasionally give rise to the flux loss phenomena at the large period limits.
We have confirmed this fact in some cases through an analytic manner.

\section{Conclusion and discussions}

In this paper, we have considered the aspects of the Jackiw-Pi vortices on a torus and, in particular, \revise{examined} an analytical calculation method to determine the vortex number \revise{in detail.}
The Jackiw-Pi equation can be derived from the Abelian Higgs model on $\mathbb{R}^2$ with critical coupling constants, which is also one of the integrable vortex equations proposed by \cite{Manton:2016waw}.
The equation can also be regarded as the ``(anti-)self-dual" equation to the Chern-Simons-Higgs theory in $2+1$ dimensions.
A meromorphic function characterizes the vortex solution to the Jackiw-Pi equation, and the first Chern number of the gauge field is interpreted as the vortex number.
If one chooses an elliptic function as the meromorphic function then the Jackiw-Pi vortex can be thought of as defined on a torus.
To determine the vortex number \revise{analytically,} we \revise{employed} a calculation method using the expansion of the Higgs field around its zeroes.
We advocate that the continuity of the Higgs fields is crucially important for determining the vortex number on a torus.
The continuity strongly restricts the fundamental domain of the vortices.

We have shown that the vortex number \revise{of the Jackiw-Pi vortex on a torus} is given 
\revise{in terms of the doubly-periodic function $f(z)$.}
\revise{
Namely, the degree $n$ zeroes of $f'(z)$ give a vortex number $n$, while the degree $n$ poles of $f$ give $n-1$, as shown in \eqref{eq:vortex_number_from_n_zeros} and \eqref{eq:vortex number from n-th pole}, respectively.}
It will provide a general procedure for computing the vortex number on a compact surface.

We have also examined some concrete examples of the vortex on a torus and its cylinder or planar limits.
Here we discuss the facet of the vortices with the ``flux loss" at such decompactification limits.
An elliptic function determines a certain elliptic curve, i.e., an algebraic curve of third or fourth order.
For example, the Weierstrass $\wp$-function parametrizes the third order elliptic curve $Y^2-4X^3+g_2X+g_3=0$, where $X=\wp(z)$ and $Y=\wp'(z)$.  
In the planar limit of the torus considered in the last section, the $\wp$-function degenerates into the rational function $1/z^2$.
This means that the elliptic curve degenerates into the third-order curve $Y^2-\revise{4}X^3=0$, which is singular at the origin.
The situation is similar to the flux loss case $f=\wp'(z)/\wp(z)$, which degenerates into $-1/z$ and the degenerate curve is $Y^2-X^4=0$.
The difference between them is that the latter is reducible or factorizable, namely, $Y^2-X^4=(Y-X^2)(Y+X^2)=0$, while the former is irreducible.
Similarly, the cylinder limits of the vortex from the Jacobi elliptic function cases share this characteristic.
The meromorphic function $f(z)=\sn(z)$ parametrizes the fourth order elliptic curve $Y^2-(1-X^2)(1-k^2X^2)=0$, where $X=\sn(z)$ and $Y=\sn'(z)=\cn(z)\dn(z)$.
The trigonometric function limit $k\to0$ reduces the curve into a quadratic curve $Y^2-1+X^2=0$, while the hyperbolic function limit $k\to1$ does into $Y^2-(1-X^2)^2=0$.
The latter is reducible to $(Y-1+X^2)(Y+1-X^2)=0$.
From the discussion in the last section, we notice the trigonometric function limit does not induce the flux loss, while the hyperbolic function limit induces it.
As can be inferred from these observations of the decompactification limits, a conjecture might be possible that the flux loss phenomena occur if the corresponding algebraic curve is factorizable into lower-order curves.
Whether this conjecture holds or is rejected, we expect that further study unvails the true mechanism of the flux loss.

Finally, we comment on the solutions with twisted periodic conditions mentioned in Introduction.
The vortex solutions considered in this paper have strict periodicity so the vortex numbers are integers.
However, there could be the vortices on a torus with twisted periodicity, which could have fractional vortex numbers as in the cases of the fractional instantons \cite{,vanBaal:1982ag,tHooft:1981nnx,Bruckmann:2007zh,Gonzalez-Arroyo:2019wpu}.
We consider such fascinating objects a subject of future research.

\section*{Acknowledgement}
The authors are grateful to the anonymous referees for their helpful comments on this paper.
K~M was supported by the Sasakawa Scientific Research Grant from The Japan Science Society.
A~N was supported in part by JSPS KAKENHI Grant Number JP 23K02794.

\newpage

\bibliographystyle{JHEP}
\bibliography{bibliography/jp-vor.bib}

\end{document}